# Towards Augmented Reality-driven Human-City Interaction: Current Research on Mobile Headsets and Future Challenges

LIK-HANG LEE*, KAIST, Republic of Korea

TRISTAN BRAUD, Division of Integrative Systems and Design (ISD), The Hong Kong University of Science and Technology, Hong Kong

SIMO HOSIO, Center for Ubiquitous Computing, The University of Oulu, Finland

PAN HUI, Department of Computer Science and Engineering, The Hong Kong University of Science and Technology, Hong Kong and Department of Computer Science, The University of Helsinki, Finland

Interaction design for Augmented Reality (AR) is gaining increasing attention from both academia and industry. This survey discusses 260 articles (68.8% of articles published between 2015 – 2019) to review the field of human interaction in connected cities with emphasis on augmented reality-driven interaction. We provide an overview of Human-City Interaction and related technological approaches, followed by reviewing the latest trends of information visualization, constrained interfaces, and embodied interaction for AR headsets. We highlight under-explored issues in interface design and input techniques that warrant further research, and conjecture that AR with complementary Conversational User Interfaces (CUIs) is a crucial enabler for ubiquitous interaction with immersive systems in smart cities. Our work helps researchers understand the current potential and future needs of AR in Human-City Interaction.

## 1 INTRODUCTION

Over the past decades, cities have evolved from concrete-and-steel infrastructures to cyber-physical entities. At first, *smart cities* [194, 201] were considered as as urban bodies leveraging communication technologies to enable service delivery and electronic data exchanges among citizens. Since then, the smart city vision has evolved toward a more technology-laden one: numerous mobile devices and Internet of Things (IoT) components interconnect and serve as critical components for human users to interact with digital entities through networked touch or gesture interfaces and conversational agents. Such ubiquity allows us to consider the smart city as a

---

*This is the corresponding author: likhang.lee@kaist.ac.kr; The first draft of this survey article was completed while the author had a research affiliation at the University of Oulu in Finland.

Authors' addresses: Lik-Hang Lee KAIST, Republic of Korea; Tristan Braud Division of Integrative Systems and Design (ISD), The Hong Kong University of Science and Technology, Hong Kong; Simo Hosio Center for Ubiquitous Computing, The University of Oulu, Finland; Pan Hui Department of Computer Science and Engineering, The Hong Kong University of Science and Technology, Hong Kong Department of Computer Science, The University of Helsinki, Finland.

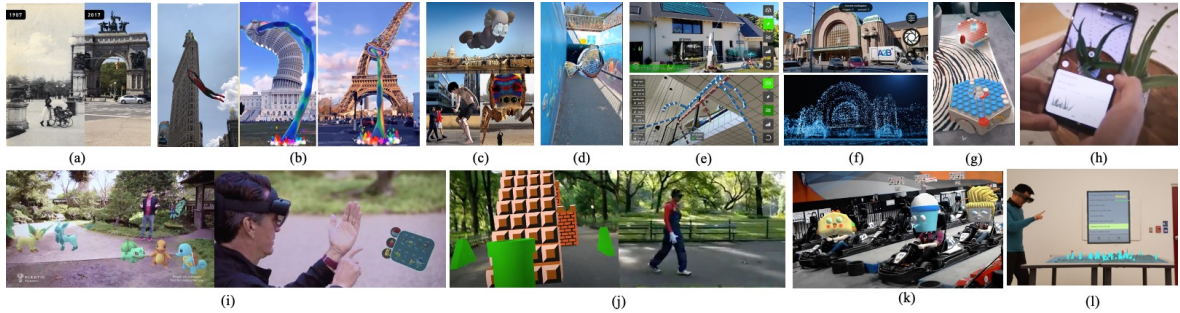

Fig. 1. Augmented Reality (AR) re-shapes our cities with real-life applications on smartphones (a − h) or AR smart-glasses/headsets (i − l). a) Urban Archive [241] shows the historical scenes of landmarks in cities; b) Snap Inc. offers AR lens to transform city landmarks into animated and amused media [242]; c) superimposing brands (e.g., Kaws) on top of the cityscape [243]; d) Turning static objects on wall arts into lively and interactive objects (e.g., fish) [244]; e) visualising the internal structure and pipelines inside an architecture, primarily for construction planning [249]; f) recording the point cloud of the building surface (e.g., Helsinki Train Station), to be reserved for adding AR objects on the details of the building [248] (i.e., precise surface / corner of a building); g) Converting an indoor floor into an interactive board game [250]; h) Scanning a real-life object (e.g., a green plant) and automatically get the related sale channels [251]; i) Pokémon Go on smartglasses named Microsoft Hololens, redefining our spatial area with virtual contents, and how the citizens interact with them (e.g., hand gestural selection on a virtual menu, supported by Microsoft MTRK) [245]; j) Augmenting a pedestrian by allocating gamified elements of Super Mario Game [246], controlled by user movements (i.e., IMU sensors); k) Adding AR features to enrich a sport in real-life (e.g., car racing with Mario Kart) [247]; l) AR Visualisation of city data collected IoT devices in the city of Toronto [252].

single computing system, where sensors and computing units are massively distributed to manage Human-City Interaction.

One timely research topic in the smart city vision is facilitating the interaction between the city and its citizens via Augmented Reality (AR) [1, 10, 194, 195], enabling citizens to access various smart city services conveniently, e.g., through wearable computers [190]. To this end, wearable AR headsets and smartglasses are enablers for user interaction with the city-system. These headsets overlay digital contents in the form of windows, icons, or more complex 3D objects on top of the physical urban environment [7]. With such headsets citizens can conveniently interact with various AR applications at their arm length, including transportation carriers, e-government services, entertainment, commerce in shopping malls, and searching for specific points of interest such as WiFi services points in AR views [173]. As shown in Figure 1, the advent of AR applications can lead to a fundamental change in how the citizens associate with our cities. In other words, AR enriches the physical world and opens new forms of user affordance [188], with the following examples. The existing AR applications on smartphones allow users to view landmarks of cities with historical developments (Figure 1a) or turn cityscape into playgrounds (Figure 1b−d). The AR digital overlay also enables the visualization of a building's inner structure to facilitate planning and design processes (Figure 1e). Another prominent feature of AR is that AR utilizes sensing technology to understand the physical world and enable user response in digital environments. For instance, scanned objects can be promptly searched on e-commerce platforms (Figure 1h). On the other hand, users with AR smartglasses/headsets are situated in a more immersive environment than the hand-held AR devices (smartphones). Although nowadays AR applications on AR smartglasses/headsets consist of trial uses of gaming, amusement, and sports (Figure 1i−k), there display a great potential for digital entities highly merged with physical environments. Practical applications on AR smartglasses/headsets can provide an overview of the urban environment to enhance the users' awareness of the city's daily changes. For instance, IoT-city data can be overlaid over digital twin representations for a higher geographical awareness when representing data in an office setting (Figure 1l).

Like any other emerging technology, the initial iterations of AR are subject to performance issues, user experience, and acceptance issues and raise multiple questions related to interaction design [1, 2]. A multitude of interfaces and interaction methods have been proposed for user interactions with city-systems (e.g., [5, 6, 8, 9]). However, the design dimensions of these interfaces and interaction methods have not been systematically discussed. Therefore, this survey article looks back, with an emphasis on recent years' developments, and synthesizes what we know about user interaction design in augmented reality (specifically focused on AR smartglasses and headsets, a specific AR form (more details are available in Section 1.1)) for city-systems and outlines key challenges for seamless and user interaction in city-system scenarios. We also strive to move beyond the individual design of user interaction prototypes toward insights on major research opportunities. The contributions of the article are as follows.

(1) provide a survey of user interaction design research on AR smartglasses and headsets in city-system scenarios,
(2) identify gaps and opportunities in the existing literature,
(3) propose a research agenda for future Human-City Interaction concerning user interaction design on AR smartglasses and headsets.

Among our main calls to action in the agenda are investigating the feasibility of AR interfaces on a city scale, developing high-speed networking for tiny form-factor user interfaces, and advancing mobile user interaction methods.

## 1.1 Preamble: Augmented Reality and Human-City Interaction

Both AR and Human-city interaction are vast domains spreading over multiple fields, including computer science and engineering, electrical engineering, social sciences, and humanities. It is, therefore, necessary to clearly define these terms in the scope of this survey.

*Augmented Reality.* Augmented reality refers to enhancing real-world environments with computer-generated virtual content through various perceptual information channels (e.g., audio, visuals, and haptics). AR brings a novel form of user experience, and user interface with our physical surrounding [231, 232]. Typical works demonstrate user-system frameworks of managing and manipulating digital overlays superimposed on the top of our physical world. In a very early study (the early 1990s), a see-through display has been mounted on the user's head, and the user in a sedentary posture can see texts and 2D menus on a workbench [223].

AR raises multiple challenges, both from a user and a technical perspective. On the one hand, the digital overlaid in front of user's views raises the need for seamless and lightweight user interaction with such overlaid [232]. A freehand interaction technique named Voodoo Dolls [230] allows users to employ two hands to select and manipulate the virtual contents with pinch gestures. Another interaction technique, namely HOMER [229], employs the visual clues of virtual hand extension as the ray-casting reaches to virtual objects. On the other hand, AR overlays have to go mobile and merge with our real-world environments, such as annotating buildings on the streets and translating texts on signage. When our urban environment serves as an enormous 3D interface, the AR elements have to be displayed in plain and palpable ways with the AR objects on top of urban elements. Significant efforts have addressed the registration errors (e.g., detection and tracking) to ensure the virtual contents displayed in the correct position relative to the real environment [225–228]. Touring Machine is regarded as the first AR prototype in mobile urban scenarios (i.e., Mobile AR). The prototype is composed of a computer and a GPS unit on the backpack and a head-worn display for showing map navigation information, and the user holds a hand-held display and its stylus for user interaction [224].

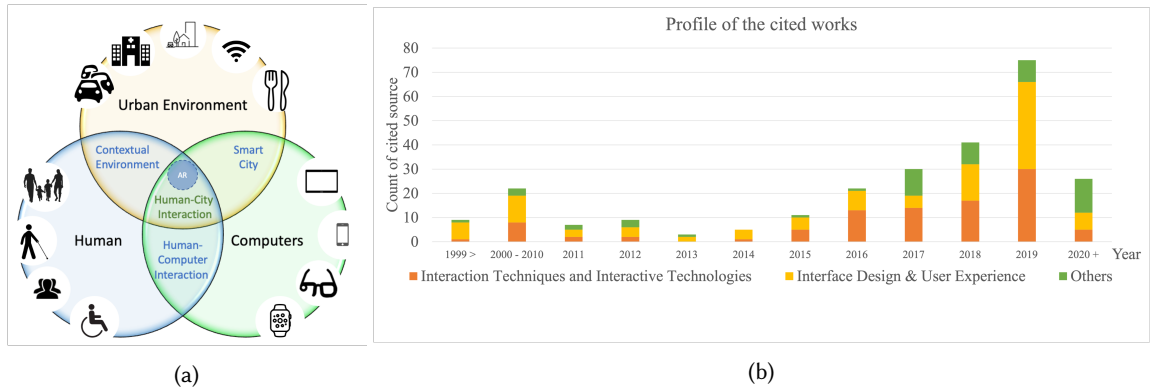

Fig. 2. (a) Human-City Interaction at the intersection of computers, human and urban environment, where AR relates to the three domains and is a key factor to drive the human-city interaction. Computers: Various mobile computing devices such as smartwatches, smartphones, smartglasses (main focus on this paper), as well as tables are candidate user interfaces; Human: Multitudinous stakeholders among the citizens including people of different races, disability, kids, and elderly; Urban Environment: city-system services such as accommodation, hospitals, traffic conditions, restaurants, and WiFi service points. (b) The counts of cited papers and online resources that are highly relevant to city-wide user interaction design for AR, sorted by interaction techniques, interface design and user experience, and other AR interaction design works.

After more than 30 years of Mobile AR development, current AR applications can be classified according to the device they are executed on. These devices include tabletops [234], ceiling projectors [235], Pico (wearable) projectors [237], hand-held touchscreen devices (e.g., smartphones) [238, 239] and AR smartglasses/ headsets [1, 2]. Tabletop devices and ceiling projectors are fixed devices that limit mobile usage. Pico projectors do not consider the user privacy of projected contents in shared space, and hence it is mainly employed for content sharing with bystander [236]. Therefore, only hand-held devices and AR smartglasses are considered as appropriate candidate devices for highly mobile AR. Although smartphones have become the mainstream testbed for AR applications, with the most remarkable example of Pokémon Go has over 1 billion downloads, two fundamental limitations (dual views and busy hand(s)) exist. First, the user needs to switch his attention between the digital contents on the touchscreen and the physical objects in real-world environments [7, 144, 238]. Second, the user's hands are occupied by holding a smartphone, and the user has to lift his hand with the hand-held device to view and interact with the AR objects. In contrast, AR headsets overcome these issues. The headset configuration allows digital overlays to be projected in front of the user's eyes and frees the user's hands from holding the devices. Such advantages facilitate mobile AR, and the users with AR smartglasses are able to experience 'the world as the user interfaces' [231].

*Human-City Interaction.* Data analytics through various sensors in the city enables the human-infrastructure-technology interactions within the urban area across the intersection of reality-virtuality [194]. The new data layer in cities allows for a multitude of novel applications. For instance, citizens can access information about traffic, shops, and construction works to navigate in the city. City authorities can assess the infrastructure's status and quickly react to critical events (e.g., traffic jam, water pipes or electric wires rupture, dysfunction in public amenities). Private companies may use pedestrian flows to optimize their store or ad campaign placements. Governing entities can use the aggregate data for policy-making to improve the city's operation. Smart cities thus demand synergy from diverse actors and benefits from different types of data access, ranging from retrospective aggregates to real-time streams.

In the connected city, augmented reality can provide context-aware interaction capacities to its dwellers, city workers, private companies, and local governments alike. Figure 2a depicts the relationship between humans (citizens), computers (e.g., wearables such as smartglasses), and the urban environment. Human-City interaction stands at the intersection of these three paradigms. Human-City interaction further extends HCI concepts related to specific computing devices' usability and design space towards user-centric interfaces for city-wide systems in a smart city. For the sake of the city-wide interaction and the high level of user mobility [232], AR smartglasses/headsets are considered as excellent candidates to deliver immersive urban interfaces. Therefore, the user interaction design follows this paradigm and moves towards the wearable head-worn computers [1]. User interaction techniques further shrink down to some smaller physical forms with new materials [128], such as smart rings and e-skin addendum on the user's body [97]. Instead of employing the sedentary mouse-and-keyboard duo, the recent advances emphasizes on-body interaction in search of subtle and convenient input solutions; for instance, barehanded pointing on GUIs using fingertips [3], text entry within the finger space [4] and even inside the miniature fingernail space [5].

## 1.2 Methodology and Related Review Articles

This survey article presents findings of a systematic literature review on augmented reality in urban environments. We reviewed a sample of 260 articles and primarily focus on works published between 2015 and 2019 (five years, 68.8%), as follows: 2020 or later: 26 (10.0%); 2019: 75 (28.8%); 2018: 41 (15.8%); 2017: 30 (11.5%); 2016: 22 (8.5%); 2015: 11 (4.2%); 2014: 5 (1.9%); 2013: 3 (1.2%); 2012: 9 (3.5%); 2011: 7 (2.7%); 2000 − 2010: 22 (8.5%); 1999 and beforehand: 9 (3.5%). The 260 sources originate from well-recognized venues of human-computer interaction as well as pervasive and ubiquitous computing, and are categorized by *interaction (input) techniques and interactive technologies* (37.7%), *interface design and user experience* (41.2%), and *others* including academic surveys, textbooks and online mini-surveys (21.2%). Figure 2b details the profile of the cited works in this article.

We found the articles primarily through publication databases such as ACM Digital Library, IEEE Xplore, ScienceDirect, and Springer Link. We used the following keywords *augmented Reality (AR), smart glasses, user interaction, mobile augmented Reality (MAR), conversational user interfaces (CUI), wearable devices, user interfaces, head-mounted displays (HMD), user attention and interruption, user vision, one-handed, same-handed, miniature design, ring-form devices, user gestures, on-skin input, smart city, urban computing, gaze pointing and selection, text entry, keyboards, embodied interaction, reality-based interaction, post-WIMP*, and combinations of these keywords. Additionally, online resources were directly searched through the Google search engine. We include papers and online resources where the user interfaces and interaction methods consider user interaction design in city-wide level or urban computing, characterized by the user contexts in physical environments as well as miniature, highly mobile, and subtle interaction. In contrast, the works on the user interfaces and interaction methods only for sedentary usages are excluded, which violates the primary baseline of highly mobile user interaction in the city-wide level. We screened through the titles and the keywords of the candidate papers and included only full papers, notes, and extended abstracts. Workshops, thesis works, talks, patents, and technical reports were excluded. When the title and keywords did not give an apparent reason for exclusion, we read the whole publication to evaluate it for inclusion. After reading the articles, we decide to include the articles based on the relevance of the user interaction with highly mobile yet immersive (mainly AR) user interaction and the application of smart cities. This leads to an initial collection of 205 articles for the review. Later on, a small addition of 40 articles and 15 online resources is included to refine this survey, leading to a total of 260 articles in this survey.

Various other surveys further informed the selection of our scope, as follows: Category I (Smart urban entities): smaller-scale interactive environments (e.g., smart homes) and IoT devices [173, 174], human-building interaction [195], and material engineering for smart interfaces [128, 129]; Category II (Wearable computing):

wearable technologies in general [176], and hardware configurations of smart wearable devices [185]; Category III (Augmented reality): mobile augmented reality and high speed networking [163], web-based AR infrastructure [163], gestural interaction with augmented reality smartglasses [1, 178], user experience measurements for AR tech [177]; and Others: conversational agents on smartphones [116].

The surveys related to smart urban entities (Category I) mainly focus on the technological aspects of building intelligent and interactive urban environments, at various levels, from smart materials and shape-changing structures to homes and buildings. The prior works on wearable computing (Category II) primarily describe the development of wearable computers and their corresponding hardware configurations and sensing technologies. However, the works from categories I and II primarily focus on the technological aspects and lack the user-centric considerations with the respective technologies. This survey connects the users through AR smartglasses (a type of wearable computers) to the smart urban entities. On the other hand, the existing surveys on augmented reality (Category III) have limited coverage on the user inputs through natural user interfaces including hand gestures and on-face/on-ear/on-forearm/on-belt/on-shoe inputs, as well as a metric collection for AR user experiences. In contrast, this survey provides a more comprehensive view of user interaction design with augmented reality for city-wide interaction, including both interface design and input techniques. The collected articles related to input techniques serve as an update to the latest development of the input techniques with AR and reinforce the trend of mobile user input towards subtle user interactions on minimal interfaces.

This survey delves into city-wide interaction issues with the emerging AR-based approaches, with the emphasis on MAR and hence AR smartglasses. We note that with the breadth of work done on related fields, this article naturally cannot provide an all-inclusive account of existing individual articles but instead focuses on the pivotal areas of state-of-the-art Human-City interaction.

### 1.3 Scope and Structure of the survey

Although AR can be achieved by various types of devices (Section 1.1), this survey mainly addresses the user interaction design on AR smartglasses that fits the purpose of city-wide interaction. This scope reflects a timely design issue on the AR smartglasses (e.g., input techniques and output content management), where the global technology giants (e.g., Apple and Samsung) are betting on AR's future on smartglasses [240] that interact with various objects in our urban life. This article starts with the human-related characteristics of Augmented Reality Interfaces (Output), such as the field of view and human vision, context-awareness, cognitive abilities, and social factors and interruptions, in Section 2. Section 3 details the paradigm shift of interaction methods (Input), including key constraints on head-worn computers, emerging hybrid interfaces, epidermal interfaces, and conversational user interfaces. Accordingly, the user-centric AR interfaces and interaction methods pave a path towards discussing Human-City interaction from the angles of city-wide interfaces and interaction methods in the smart urban environment. Following these, we revisit the field to lay out a research agenda (in Section 4) that will help researchers working on AR and smart cities to contextualize and focus their efforts.

## 2 AUGMENTED REALITY INTERFACES

In the mobile city-wide urban scenarios, users obtain visual information through the see-thru display on AR smartglasses. AR paves the way to transform how we interact with the digital entities of the smart city. By blending the virtual world with the physical worlds, AR enables numerous applications through well-designed interfaces [175]. Such applications range from government services and in-situ architecture to interacting with smart IoT devices. Considering city-wide applications, the smartglasses should display the information effectively anytime and anywhere, for instance, checking text messages in walking posture or browsing social networking images during daily commutes. In such circumstances, user motion shakes the smartglasses' display due to the unavoidable vibration from walking and commuting, hence impacting readability [12]. Also, users read the information on the see-thru display with diverse and unpredictable backgrounds. The user's attention will be drawn to the surrounding physical objects and virtual objects simultaneously [17]. The information display often swings between the choices of human central and peripheral vision. If the digital overlays occupy the central vision on the see-thru display, the interaction between the users and their physical surroundings in the city will be interrupted. Therefore, peripheral vision [7, 16] becomes an alternative to maintain the user multi-tasking ability [18] in urban scenarios. Apart from the conventional displays primarily concerning the combination of font type and size [24], additional dimensions of information display, including environmental effects to the visual ability [19], background management [187], user mobility [21], the timing of notification [20], should be further considered. We classify the common issues of AR interfaces into six categories, and Table 1 (Appendix) summarizes the most recent works, with the following brief explanations: 1) *Cognition* refers to the user cognitive loads in dual-task situations that lead the AR user to switch his/her attention between the real world and digital environments. 2) *Content Access* focuses on managing and displaying the AR contents for reduced interaction cost and hence improved usability. 3) *Field of view* indicates the design constraints of limited display size on AR smartglasses. 4) *Human vision* considers the limitations of the human visual functions, peripheral visual in particular, during the design of content display in AR. 5) *Readability* means the user's ability to read or receive the contents with various forms and sizes in different environments and conditions (e.g., illumination) inside our cities. 6) *Social* concerns about the appropriate timing of interruptions matched to the physical environment or the social factors raised by bystanders.

## 2.1 Field of View and Human Vision

The Field of View (FOV) of a mobile headset directly determines the digital object's size and position overlaid in front of the user's sight. Most commercially available AR headsets present a FOV smaller than 60 degrees (Figure 3). Such a FOV is far narrower than the typical FOV of a user with 10/10 vision. For example, Microsoft Hololens (gen. 1) provides a 30 X 17-degree FOV, implying a 34.5-degree FOV diagonally in the screen ratio of 16:9. This FOV is similar to a 15-inch diagonal, 16:9 screen located 2 feet away from the user. Other smart glasses such as DAQRI or Meta 2 present a FOV equivalent to a screen of the size of 1x − 3x a 9.7 inch iPad Pro (240 mm (9.4 in) (h) 169.5 mm (6.67 in)(w)) [162] in an arm length (approximate to 800 mm). In contrast, low-end smartglasses present an even smaller FOV. For instance, Google Glass has a FOV equivalent to a 25-inch screen from 8 feet away. It can barely accommodate one-sentence messages and notifications. Most apps should thus keep the information simple and design the interfaces wisely. The huge FOV gap between the smartglasses and the user sight negatively impacts the user's experience and deteriorates task performance [7]. Prior works show evidence that restricting a person's FOV down to less than 50°on head-worn displays can noticeably hurt the usability and hence the task performance [26].

The peripheral visual field is an essential part of the human vision and is helpful for daily locomotive activities such as walking, driving, and sports [29]. Visual cues from the periphery help to detect obstacles, avoid accidents, and ensure proper foot placement while walking. Smartglasses and AR headsets with limited FOV may exploit such property in their interface design as they partially cover the user's peripheral field. Apart from leveraging non-visual cues such as audio and vibrotactile feedback [167], several recent studies [2, 7, 16] employ peripheral vision to display information a edge areas inside the smartglasses' displays to alleviate the issue of limited FOV. In an AR navigation application [7], the notification information appears at the peripheral visual field. The authors exploit the peripheral visual field's high sensitivity to motion to provide navigational information to the users without distracting them from their main task, i.e., the users maintain their attention to the main task with their central vision. With that concept in mind, they design a simplified navigation application providing three instructions: straight, left, or right. Their experiments under various backgrounds and lighting conditions show that most users reached their destination successfully while looking at the screen 50% less than for a traditional navigation application. A prototype of AR contact lens display [159], although only providing a single-pixel visual cue, can leverage the peripheral visual field for the above navigation tasks, i.e., four separate pixels in the top, down, left, right position represent the direction. The activation of one pixel serves as the movement and direction clues. AR contact lens is moving from laboratory to the market with more pixels and better resolution [160], and we see a blank slate for designing various applications in the urban environment with AR contact lens and peripheral vision altogether. It is important to note that bulky spectacle frames (e.g., Microsoft Hololens) occlude the peripheral visual field, so users lose awareness in the occluded regions of interest and behave less responsively to critical situations [126]. In contrast, contact lens get rid of the headset occlusion and open exciting possibilities in the peripheral visual field.

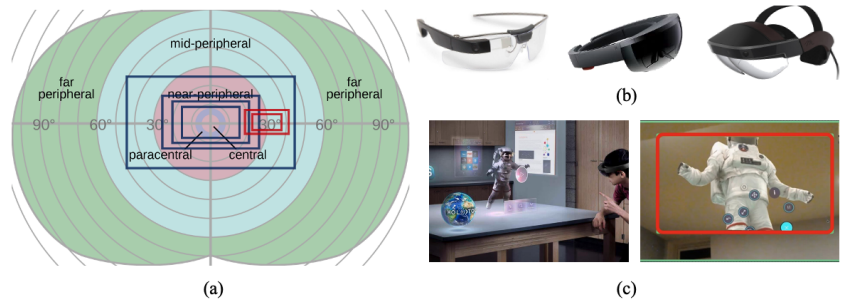

Fig. 3. a) Field of View (FOV) of Human Vision, in which the space between one circle edge to another edge is equal to 10°. E.g., the 88° FOV of Meta consists of 44°–(left) and 44°–(right) from the central vision. The FOV of AR headsets, represented by squares inwards to outwards: Google Glass (red, 15°), Espon BT300 (red, 23°), Hololens (blue, 30°), DAQRI (blue, 40°), Atheer AiR (blue, 50°), Meta (blue, 88°); 5 out of 6 have less than 60-degree FOV, which is significantly narrower than the human vision of 220°; b) Examples of AR Headsets (From Left to Right): Google Glass, Microsoft Hololens, and MEta; c) The comparison between the projected AR scene in a marketing campaign (left-hand side) and the actual FOV viewed by the user with Microsoft Hololens, showing a partial body of an astronaut (right-hand side).

Besides peripheral vision, the user vision is subject to multiple factors affecting the content readability and legibility, such as color coding and illumination [49], size and style [24], shakiness due to body movement [12], visual discomfort and fatigue [169], and as well as the content placement [21]. Although many information presentation methods [13] and automatic systems for optimal content placement have been proposed [28], the existing works primarily limit their studies on evaluating a single factor and put their attention principally on the textual contents. The above studies are important to the use-cases in urban environments. Such research studies are also extended to other topics such as visuals and graphics for data presentation. Considering that the users may interact with smartglasses in city-wide applications, graphics and charts proactively act as an auxiliary tool for decision-making. Smartglasses users can easily read data charts in a laboratory or office environment. However, these charts become difficult-to-see under powerful illumination in outdoor or mobile scenarios. The users will also be influenced by the Ebbinghaus illusion [15] when a bar chart is overlaid on the see-thru screen over a background presenting square-shaped patterns. It is necessary to adjust the chart presentation according to the background for better readability in such a scenario. However, limited research efforts address optical illusions for see-thru displays, such as virtual objects size [50] and data presentation due to Ebbinghaus illusion [15].

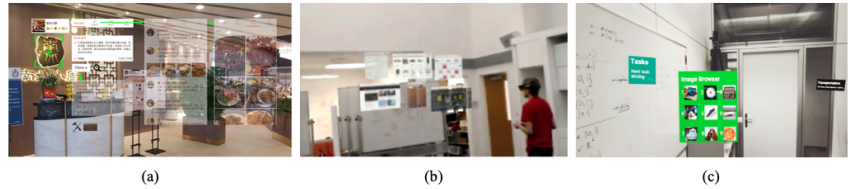

Fig. 4. Context-aware framework examples considering the user's contexts: a) food menus and reviews in a restaurant [6], mainly based on the user's geographical location; b) placement of AR content with multiple windows in a classroom considering the user's movements and physical surface [22]; c) showing AR contents considering user's in-situ task nature (e.g., task workload) [23].

## 2.2 Context-aware Interface Design

Interaction in AR is a delicate issue. Interface design strategies borrowed from the desktop computer world are often unpractical. Such concerns become predominant when using AR for a typical desktop/mobile application such as web browsing. If we consider the smartphone world, it quickly became apparent that websites had to be adapted to the small screens. Then, to accommodate the increasing number of screen sizes and resolutions with various pervasive display [161], the development of the multiple versions of the website got incorporated into the main website's design workflow. M2A framework [6] aims at replicating this process for AR. However, AR's specificities force us to ask the following questions: 1) What is the optimal visualization paradigm for AR websites? 2) What is the optimal interaction paradigm for AR websites? 3) How to practically design a website for AR? The two first points are closely interleaved: user interaction is often impractical. On the other hand, the AR virtual world is 3D, virtually unlimited, and allows to display much more content than a traditional screen. We can dramatically limit the number of interactions by flattening the website's structure and presenting it at once in the AR environment. That is, the simplified information will lower the cost of user interaction in terms of click and scroll [146], and lead to more engaging and meaningful interaction in the appropriate contexts [149]. Furthermore, to avoid inputting the URL, we can use AR's context-awareness to automatically display the webpage when and where it should be displayed. In M2A [6], the webpage of OpenRice (a popular restaurant rating website) on top of the restaurant's facade serves as an example of context-aware interaction with digital entities in our cities. Users can quickly locate the relevant information in a timely, intuitive, and convenient manner. Regarding the practical implementation of an AR website, the M2A engine automatically extracts the main elements of a webpage and renders them in AR. M2A also allows web developers to refine the placement of the AR blocks on top of the physical world.

M2A serves as a fundamental example of reorganizing interfaces from volumetric interfaces designated for the desktop environment to lean and accurate information suitable for the neighboring urban situation. The level of details significantly impacts the user affordance [188], mainly caused by the high cognitive load within the small FOV on smartglasses in mobile scenarios. We illustrate an example here. While a long and demanding (high cognitive load) text appears on the limited FOV display, the users only need small visual cues corresponding to the immediate task. One may argue that users can switch to other devices of appropriate capacity or temporarily suspend the task. However, switching between tasks and devices would impose a suffocating blockade on ubiquitous city-wide interaction. In this sense, the context-aware information display emphasizes the consideration of the tasks and applications the users are dealing with as well as the user situations. An optimized mapping between the task/application nature and the level of information details can bring users prompt information at the appropriate level of cognitive load. A rule-based and integer-LP optimizer [23] maintains a context-aware AR interface adapting to the user's mental workload in one specific task or environment. In a city-wide implementation, the number of AR contents and the related interface designs will require new capacities that cannot be met by redeploying existing content creators and their manual capacity. City-wide AR will thus require new computational approaches like the framework and optimizer mentioned above.

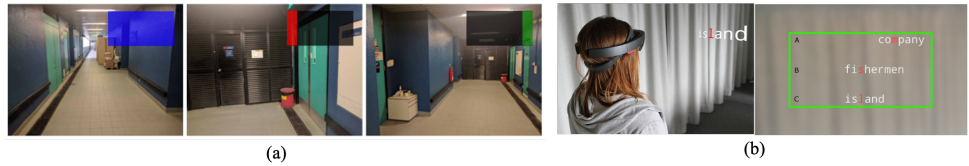

(a)                                                                          (b)

Fig. 5. Examples of visual cues that alleviate the user's cognitive loads a) direction cues (blue – go straight, red – left, green – right) in a navigation task in a building [7], b) investigating the text display pattern when the user simultaneously moves forward [21].

Context-aware information organization on the AR smartglasses display is an integral part of the problem but should not be an end in itself. Instead, the broad adoption of new display paradigms will succeed only as far as they consider the content projected on the surrounding physical objects in the urban environment and the user-body movement. A recent work on design and usage patterns for AR [22] states that the content display in the physical environment is subject to three metrics: user movements, physical surfaces, and the relationship between multiple windows. Such metrics imply a mutual consideration among compatibility with real-world activities, vision-body coordination, and user performance for output overlaid on the blend interfaces of the digital and physical world. Although the high-end smartglasses (e.g., Microsoft Hololens) can describe the physical environment through obtaining the spatial mesh data, the speed of scanning such mesh data becomes a bottleneck. It thus cannot catch up with the user's walking pace in city-wide interaction. One of the possible solutions resides in caching the city physical surfaces as re-usable and ready-to-call spatial anchors when the user arrives at a recognizable location [22].

User interface design serves as a critical factor for a better user experience through AR with the smart city. Figure 4 shows the examples of AR content management in our urban environments such as restaurants, classrooms, and office areas. In summary, the most recent works on AR interfaces become more intelligent and interactive through context-aware architecture. The interfaces with corresponding interaction techniques can significantly contribute to productivity, efficiency, and naturalness of Human-City Interaction. As certain intelligent interface paradigms and techniques are directly conflicting, there is a need for a comprehensive architecture acting as a "mediator." Throughout the above three examples, we illustrate key interface patterns in which the context-aware interfaces for major AR scenarios should support.

## 2.3 Cognitive Ability

Besides the context-aware interfaces fueling diverse city-wide applications and services, the user perception of such interfaces is another essential facet of the interaction between human users and our cities. The original intention of AR with see-thru optical display is to blend the virtual and physical environment into an integrated interface e.g., road navigation in a building and reading texts while walking (Figure 5). Both the digital and physical information are perceived by the users simultaneously. Well-known industrial applications of AR smartglasses include the maintenance of complex machinery such as cars and aircraft. Instead of reading the instructions on an adjacent mobile device screen, users can receive and process the digital and physical information in parallel with head-worn displays [37]. However, user attention [144] is a scarce resource in the AR setting [18]. The digital overlays unavoidably distract the user from the physical world, causing a negative performance, e.g., overlooking details on physical items [39].

Similarly, modern cities can be seen as complex systems through which citizens, workers, and authorities may interact using Augmented Reality. Users interacting with the city systems through AR smartglasses deal with the same type of parallel information as in industrial applications. Considering that the user focuses his/her central eye gaze on the smartglasses display at a close focal point, his/her multi-tasking ability is strongly impacted. Mishandling user attention will lead to devastating effects on user performance and even safety [170]. For instance, a user driving a car on the highway at 150km/h who takes his eyes off the road for one second is blind for 42 meters. There have been about 5,984 pedestrian traffic fatalities in 2017. One of the leading causes of such accidents is the divided attention to mobile devices [40]. Hence, studies on micro-interaction for collision warning [122] in vehicular interfaces [124] are rising topics in the research community. User attention is another critical metric in designing the city-wide implementation of AR. It is crucial to take the user's cognitive limitations into account, where minimum user attention [40], least interference with the tasks in physical environments [18] and the aforementioned information overload [2] serve as the high-level design principles for daily dual-tasks in urban situations.

To the best of our knowledge, a minimal number of existing works study user performance in multi-task situations using smartglasses in a mobile scenario. Users show degraded performance in both task completion time in the physical environment (primary tasks) and their comprehension of the digital contents while sitting, walking [20], sport activities (e.g., climbing a wall [36]), vehicle driving [165, 170] and even in unexpected events (e.g., encountering a person with a wave hand [38]). Rzayev et al. [20] give design clues to the effects of the placement position of the textual digital overlays on the user's cognitive load. Their study reflects that the digital overlay at the 'bottom-center' is better than the 'center' and 'top-right' as users feel less distracted and more confident during walking, with the bonus of reserving the user attention and encouraging them to walk with increased speed. The study of [27] focuses on image content. It shows that the edge position ('middle-right', 'top-center', and 'top-right'), eventually in the peripheral vision area, are suitable for displaying image contents when the secondary stimuli are less important than the real-world task. Also, their results reflect that the 'middle-center' and 'bottom-center' positions should be reserved for dual-task scenarios requiring high noticeability and constant updates on the secondary stimuli shown on the smartglasses display. Another study [18] introduces walking tasks with obstacles simulating the real-world interference in urban environments. They further outline design implications such as 1) removal of AR content in hazardous and time-critical situations in the real world; 2) performance allowance for user information retrieval in the complicated mobile scenarios through adaptive information display speed; 3) information placement within or outside the FOV depending on the importance of information as well as the contexts whether it is user's related task or unknown usage context. An additional study of information placement among 'inside FOV', 'on-(user)body', 'Floating', and 'In-Situ' (e.g., the nearby walls) evaluates the user perception in terms of intrusiveness, noticeability, comprehensiveness, and urgency [20].

## 2.4 Interruptions and Social Factors

The AR output in the urban environment goes beyond the scope of the tasks. User attention can be interrupted by other users in such a dynamic environment. User Interruptibility is a decade-old problem in the field of HCI, and previous works illustrate the cost of interruptions [42] and the user performance degradation [41], although interruptions are the unavoidable part in teamwork and collaboration in any organization [43]. Among the existing studies, the cost of interruptions is usually quantified as the penalties with negative consequences. The well-established yet socially acceptable solution is to perform the interruptions between tasks – the opportune moment of task switching [44], which minimize the time and cognitive efforts from re-focusing on a task [45] as well as the tolerance to the number of interruptions [46]. Users with AR smartglasses will encounter similar issues in the urban environment, where surrounding bystanders (e.g., friends and colleagues) may disturb the user immersed in the tasks interacting with the city systems. Additionally, city-wide use-cases on smartglasses are very different from the office and laboratory setting, implying that isolating oneself from the real world is impossible [48]. Unlike smartphones with more discernible timing, the smartglasses display is constantly at a close distance from the user. In other words, the user's central and peripheral visions are constantly disturbed by the AR overlaid. The necessity of identifying interruption timing will steadily grow in importance over the longer term. A most recent work [47] proposes several design directions for head-mounted displays, including 1) Enhancing awareness on Interruption Location; 2) Supporting swift and accurate interruption; 3) Supporting collaboration by addressing bystanders; 4) Knowing the user's task in advance; 5) Recognizing clues through Gestures and Real-world conventions; 6) Designing non-discernible task switches.

Apart from solutions relying on social conventions, sensory technology (e.g., computer vision) in the smart city can decide whether to interrupt a smartglasses' user occupied with a task through recognizing the body gesture and head direction [45, 47]. When smartglasses become popular mobile devices as nowadays smartphones, the bystanders can employ the camera embedded on smartglasses to receive the clues of interruption opportunities [44], with the support of cloud and edge servers working on the computationally demanding tasks of gesture recognition [10]. With mobile technology advancement, bystanders can acquire information such as the user's location and task importance. Exploiting computer vision and mobile technology enables quicker and more accurate guesses of the interruption timing. Additionally, considering that interruptions are frequently happening in urban situations [42], the approaches employing peripheral vision can be a strategy reducing the harms of interruptions to the primary tasks at the central vision [19], and potentially drive the task forward. Overall speaking, the information organization and their design studies, under the premise of employing augmented reality smartglasses display, are still in their infancy stage. Moreover, plenty of research opportunities exists when the domain moves to the wild for human-city interaction.

Finally, social acceptability is a crucial factor in the wide adoption of mobile headsets [255]. It primarily evaluates whether the users are willing to use AR-driven content display in front of a certain audience or at a certain location, namely the audience-and-location axes [256]. Suppose the mobile headset users and the bystanders have adverse reactions to the usage of mobile headsets [257]. In that case, there exist less effective ways to deliver AR contents through the head-worn display. Due to privacy concerns, a user in a private location has less willingness to employ the sensor. Also, users in a conversation are cautious of the headset owners [254]. The trade-offs between user privacy and usability remains a challenging issue in content management and display. The context awareness, as discussed in Section 2.2, can impact the adaptability of user interfaces in such locations and hence the user experience. Without the available information due to the privacy restriction from the audience-and-location axes [256], M2A [6] cannot easily screen out irrelevant contents and locate the space for content display. M2A leverages geographical information (i.e., the location axis) to select candidate contents to be shown in the limit-size display. Next, M2A utilizes the camera on the headset to recognize the blank space for content display in a physical environment. However, the camera may capture certain individuals in the background, and hence potentially conflicts with the privacy of the users and the bystanders (i.e., the audience axes).

## 3 INPUT APPROACHES: A PARADIGM SHIFT

Before the mouse was invented for desktop computers, interacting with objects on the screen was clumsy and indirect [30]. The mouse made interaction with personal computers more user-friendly than ever before and led to the popularity of personal computers [31]. Similarly, touchscreen technology led to the widespread use of smartphones [32] due to its user-friendly design [33], which enables the users to manipulate the objects on the screen directly [34]. Wearable head-worn computers such as smartglasses face similar issues as early desktop computers and smartphones. Their usage is limited because of the bottleneck in user-friendliness. As stated in Section 2, AR smartglasses are a good candidate to connect users with immersive interfaces to the city-system.

However, without proper input approaches, smartglasses can only act as a one-way display. Users have no practical way to interact with the city-systems' dynamic content shown on the displays' interface. In this section, we first discuss the obstacle of smartglasses interaction and the corresponding strategy. Next, we summarize the trends of the recent research efforts and revisit the interaction challenges [1] on smartglasses. Finally, as outlined in Figure 10, conversational user interface scan potentially contribute to the human-city interaction.

## 3.1 The Constrained Head-worn Computers

The key differences between desktop computers/smartphones and wearable computers are the more constrained interfaces of head-worn computers (Figure 6), including screen real estate (i.e., limited FOV, Section 2.1), hardware configurations [3], tiny touch-based interface [2], and user mobility [4]. The detailed explanations for the latter three constraints are as follows.

*(A) Constrained hardware configuration:* Smartglasses present limited computation power and short battery life, which are not favorable for computationally intensive tasks, for instance, computer-vision-supported hand gestures for interacting with the icons and menus on smartglasses [151]. Google Glass (low-end smartglasses in the current market) contains an ARM Cortex-A9 MPCore SMP at 1 GHz and displays 1 – 3 hours of battery life. This configuration is similar to desktop computers in 2000, where Intel Inc. claimed that it was the first to market the CPU featuring a 1 GHZ clock speed [51]. Even though the semiconductor manufacturers can produce small yet powerful chipsets for smartglasses, the battery will be used up quickly if intensive computation tasks are running [185]. Running energy-consuming algorithms on smartglasses will severely hurt the sustainability as stated by Yann LeCun as the next grand challenge of machine learning [62]. Hence, the design of machine-learning-supported user interaction should carefully alleviate the constraints and facilitate the day-long usage of AR smartglasses in immersive urban environments.

*(B) Indirect touch on miniature interfaces:* It is more difficult to accurately hit small targets on a miniature-size touch interface [60]. Due to this constraint, smartglasses often only serve as an extended display to smartphones. Similar to smartwatches, their functions are limited to message notification, bio-metric information collection, user-health status monitoring, as well as location positioning and city navigation [3]. Besides, text entry for message input on smartglasses is usually restricted to predefined texts and emojis for one-click replies because of the size of the interfaces [4].

*(C) Physical constraint from mobility:* The small interfaces for text entry on nowadays smartglasses have not thoroughly considered the issues of mobility and social acceptance, on the top of the input easiness [164]. For instance, the touch interfaces on the frame of Google Glass and the mid-air hand gestures of Microsoft HoloLens require lifting the hand to the eye level, which is tedious and draws unwanted attention from the surroundings. Thumb-to-finger interaction [4] can serve as an alternative to unnoticeable and subtle text entry [52]. However, the small-size finger space can barely accommodate the full QWERTY keyboard with two hands [53]. Two-handed text entry is not suitable for the mobile situation [61], e.g., holding a shopping bag, and thus single-handed text entry becomes necessary [5].

## 3.2 Strategies for Constrained Interfaces

The body-worn wearable computers serve as an extension of our body [55]. However, the small-size wearable computers attached to human bodies pose various constraints (Section 3.1). The current design of interaction techniques for wearable computers is derived from the tangible and spacious desktop interfaces. The direct adoption of such interfaces on wearable computers makes user interaction difficult and will severely decrease adoption. For example, the gestural input on Hololens is four times slower than the tangible mouse pointing device, with diminishes user satisfaction by more than 50% [179]. O'Sullivan and Igoe [54] depict desktop computers human users as an alien with only one finger, an eye and two ears, which means that not all the capabilities of human beings are currently exploited. Perhaps the head-worn devices and the users' physical body and cognitive ability can bind together as an integrated entity of input and output, thus resolving the constraints mentioned above and establishing appropriate input capability. Knowing that human users are very skillful at the physical world using their body [56], the users can become a part of the windows, icons, menus, and pointers (WIMP) [57].

Originated from the theories of embodiment [58] that focus on how our bodies and active experiences influence how we perceive, feel and think, embodied interaction [59] advocates that the habits, skills, experiences, and abilities of human being that we already have should be at the core of designing the interaction interfaces and techniques. In other words, the users' physical body and cognitive sense act as key drivers in the user interaction experience. Head-worn wearable computers and embodied interaction can be viewed holistically as the coincidence of input and output interfaces. Embodied interaction serves as a *design strategy* to improve the input capabilities of the constrained interfaces of head-worn wearable computers (Figure 6). The high-level strategy is that wearable computers will adapt to humans through well-designed embodied interaction (Table 3, Appendix). Considering that head-worn computers serve as an extension of the human body, enriched with content from both the blended digital and physical worlds, we advocate human users' symbiosis and wearable computers and push the emerging landscape of head-worn computers towards more usable interfaces as well as bidirectional devices. By examining the aforementioned three constrained scenarios (see Section 3.1), the optimal interaction techniques between human users and wearable computers are identified. These interaction techniques (Figure 6), as explained in the next paragraph, leverage the advantages of the humans' physical body, experiences and skills [157].

Dexterous fingertips [3] can serve as a pointing device for mid-air interaction on AR smartglasses supported by a computational inexpensive algorithm (Figure 6 (1)). Additionally, hand gestures are regarded as user-accepted approaches to drag office documents in an AR workspace [9]. Second, text entry approaches leverage the human knowledge and customs such as the alphabetic order [2] (Figure 6 (2)), the writing stroke orders of roman alphabets [74], and the familiarity with the QWERTY [5, 76] and alphabetical [83] layouts for gestural typing [64], to achieve space-saving text entry interfaces in the limited FOV of head-mounted computers (Section 2.1). Third, users can distinguish the force levels and hence work on text acquisition in dense and cluttered environment [60]. Finally, human thumbs can naturally locate the keypads within the finger space [4, 53, 61] and on the fingertip [5] (Figure 6 (3 and 4)). More interaction techniques leveraging the theories of embodiment are shown in Table 3. So far, we have discussed the obstacles of interacting with smartglasses, and the strategies of embodied interaction for smartglasses. In the next paragraphs, we discuss the most recent research works and summarize the latest input interaction design trends on smartglasses.

## 3.3 Emerging Hybrid Interfaces

An earlier survey regarding smartglasses input defines four categories for classifying input techniques on smartglasses [1]. These categories are: on-device interaction [64], on-body interaction [75] [95], hands-free interaction (e.g., gaze [88] and whole-body gestures [93] [96] for emoji inputs [140]), and freehand interaction with wearable cameras [86] or sensors inside the closed environment of a vehicle [120], showing clear boundaries among the four categories. Nevertheless, as reflected in the most recent works, these boundaries are becoming more blurred. Embodied interaction can serve as a promising strategy to alleviate the highly constrained environments on AR smartglasses. The incentives of exploiting the resources available to the users themselves coincidentally drive the most recent input techniques and interfaces to hybrid and multi-modal approaches. The number of hybrid approaches is minimal, and the studies on hybrid approaches primarily focus on touch-based interaction and hand gestures [1]. Although hand gestures for manipulating virtual 3D contents on smartglasses, such as object rotation and translation by simply rotating the wrist and swiping the hand, are intuitive, natural [86] and easy-to-learn [171], employing sole hand gestural approaches suffer from long dwell times and performance degradation [3] due to the Midas problem [125]. In earlier works on hybrid approaches, the touch-based interface serves as a swift and responsive switch to distinguish the users' hand gestures from the deliberate interaction with AR overlays. In contrast, the most recent works on on-body interactions treat the touch-based interfaces far more than a switch, and the task natures are more diversified and complicated.

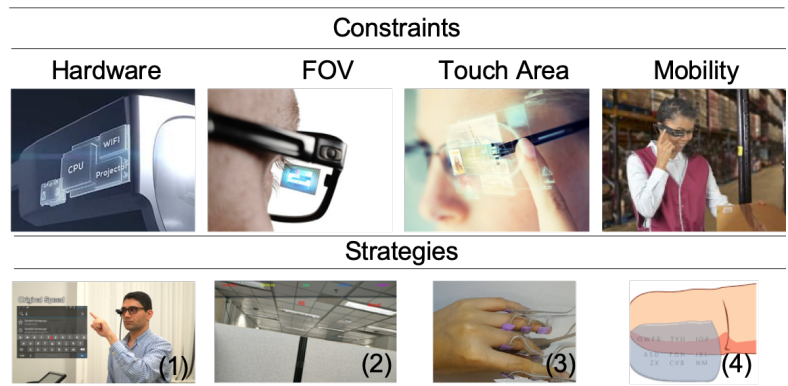

Fig. 6. Constrained Interfaces and examples of embodied strategy: (1) Energy-saving and computational inexpensive algorithms [9] allowing fingertip interactions in mid-air as a pointing and selecting device; (2) Imaginary Keyboard-less interfaces that release the FOV from the occluded view of a standard QWERTY keyboard [2]; (3) On-body touch surface as a supplementary touch area, e.g., a palm [6]; (4) Miniature-size interfaces (e.g., on a finger [5]) for highly mobile input channels.

A nail-mounted gestural input surface enables users to apply press and swipe gestures, i.e., reaching from one fingertip to the surface on another finger to interact with objects on mobile devices [80]. In TipText [5], the user can perform a thumb reach on the index finger, where an ambiguous QWERTY-like keyboard locates on the miniature touch surface on the first phalanx of the index finger. Text entry is regarded as a series of repetitive target acquisition tasks on the keypads inside a keyboard. Leveraging the human skill on thumb-to-finger interaction can ease the error-prone and tedious tasks [4]. Similarly, with FingerT9 [61] the users perform thumb-based interaction within the finger space in subtle and natural manners, enabling users to achieve text entry on touch-sensitive buttons inside the finger space. In a gaze-assisted text entry system named GAT [87], the touch-based controller can significantly improve the accuracy of pointing-and-selection.

### 3.3.1 On-body Techniques and Everyday Objects.
In this paragraph, we only consider the on-body input techniques on small-size wearable interfaces because of its high mobility nature for city-wide interaction in the urban environment (Figure 7b – c). The most fundamental way is to attach sensors in the form-factors of buttons [70] and e-skin [72]. For example, DeformWear [70] are button-sized interfaces, attachable on various body locations, that augment the user's body with more expressive and precise input capability with pressure, shear, and pinch deformations. Other on-body techniques [70], especially those on arms [90, 100] and fingers [4, 79], reserve the user's hand from occupancy at all time. Daily objects are readily available and apparent for being used as input interfaces [86]. Unlike hand-held devices, users can immediately return to the primary task in dual tasks scenarios.

Regarding interaction techniques in AR, barehanded object pointing and selection have long been considered [152]. However, with solely computer vision-based approaches, accurate and responsive recognition of hand gestures is technically challenging. For example, points and taps in mid-air require CPU-hungry applications [3] and hence lead to a significant latency, hurting the user experience [1]. Instead, IMU-driven [100], optical [79], acoustic sound [75], Pyroelectric Infrared [78], electromagnetic [147] and capacitive [90] sensing capabilities are commonly applied to augment the capability of body sensing. In [100], a vision-based system detects the hand location and an IMU-driven ring device determines the touch event with virtual overlays in AR. Below are some examples of finger-to-arm interaction. ActiTouch [90] presents two electrodes in the capacitive sensors, and touch events from a finger of one hand on the forearm of another hand close the RF signal paths in the circuit. Through a precise on-skin touch segmentation [75], the user can perform taps on the icons and menus inside the digital overlays projected on the user's arm. Similarly, on-skin touch can be further applied to interpersonal touch interactions for social applications[91, 92]. Another emerging stream of on-body interaction is the single-handed and same-hand interaction within the finger space, characterized by more subtle interaction than the finger-to-arm interaction. The formative studies on the button positions on the small space of a palm in 2D [119] and 3D [68] and a finger [5] have been conducted. FingerT9 [61] and TipText [5] are examples of input techniques within a palm and a finger (Figure 7d). The above finger-to-arm wearable solutions need a touch-sensitive surface at the user's palm to support the input procedures. These solutions share a common goal to integrate interaction solution seamlessly and even invisibly into our bodies. In a similar way, another emerging approach is to develop digital textile to integrate interactive materials and conductive threads into our fabrics (e.g., bags and clothes). As suggested by a project named Jacquard [258], manufacturing interactivity woven can become inexpensive in a large-scale manner. Users with such woven on trousers and jackets can achieve on-demand interaction with our AR urban. Two recent examples demonstrate that digital textile could become an on-demand interaction surface for inputs on mobile headsets. First, PocketThumb [259] is a wearable touch interface embedded into the fabrics of a front trouser pocket. It offers a dual-side interactive surface that enables users to do touch-based interaction (e.g., controlling a cursor) with thumb sliding along the inner side of the fabric, and simultaneously tapping or swiping with fingers on the outer side of the fabric. Second, ARCord [260] provides an addendum of cord-based textile attached to the user's body. A tangible controller at the end of the cord allows users to initiate menu selection and ray-casting object selection on mobile headsets.

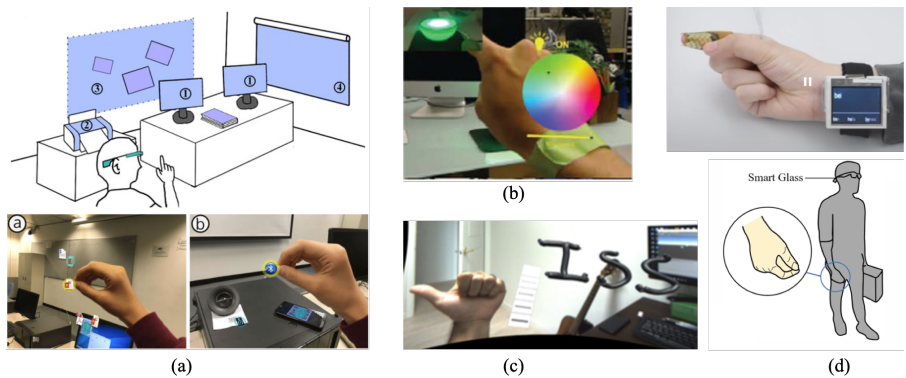

Fig. 7. Examples of On-body Interaction Techniques and Everyday Objects. a) Ubii [7] Interacting with everyday objects in office environments such as dragging PowerPoint file from a desktop to a projector screen, and turning off/on a speaker; b) An on-body menu to control the led display light by hand gestural commands [172]; c) AR authoring (writing text) with on-body menu driven by hand gestures [151]; d) a miniature input interface on fingertip area for text entry with smartglasses [2].

On the other hand, the input channels have not been restricted to wearable on-body interfaces. Everyday objects, such as smart IoT devices [172, 199], and tangible objects, in particular, become another unmissable aspect of user interaction. Hence, research on interaction techniques with tangible objects in the post-WIMP interaction era is emerging [130], as tangible objects construct an input-output relationship in our society and computer systems [145] The most recent input techniques with everyday objects includes micro-gestures interaction by holding objects [63, 134, 137], interacting with smart objects [132, 133], as well as re-configurable UIs [127]. Furthermore, a user elicitation survey investigates users' micro-gestures when the user is holding an object of various sizes and objects such as pestle, marker, needle, A4 paper, credit card, or suitcase [63]. All the object holding gestures are generalized into the six following object types: Cylindrical, Palmar, Hook, Lateral, Tip, Spherical, and the gestures are further classified into On-object, On-body, and In-air interaction. Also, as for in-vehicle systems [121], researchers design finger micro-gestures while the user's hands are holding a driving wheel. Nowaday's mini-radar such as Google Soli [84] and passive infrared sensor [78], RFID and IMUs [186] can leverage the results of micro-gesture elicitation for gestural design with recognized daily objects [137]. Aside from micro-gestures with holding objects, the research community uses daily objects as an alternative option for input devices. UnicrePaint [134] enables users to draw an apple with a real-life apple on a tangible tabletop interface. Through micro-gestures measured by a smartwatch, users can interact with the increasingly popular large display on a wall surface [132]. In [133], the building infrastructures are augmented with ultrasonic acoustic sensing capabilities, and consequently, the building surface becomes sensitive to the user's press and swipe and recognizes a number of gestures. Ubii [9] (Figure 7a) attempts to break the centralized user interface employed on desktop computers into pieces woven in the domain environment. The users interact with these AR interfaces paired to the physical objects (e.g., walls and printers in an office), where physical and digital presentations are matched in the same context. Finally, 82 configurable objects, including openable lamps, sword-canes, teapots and cups, sand and clay, etc., were studied and to formulate the corresponding design implications for re-configurable UIs [127]. The results of this study can be further applied to the design of interactive surfaces composed of shape-memory and adaptive materials, which can be reprogrammed as re-configurable, shape-changing interfaces [129].

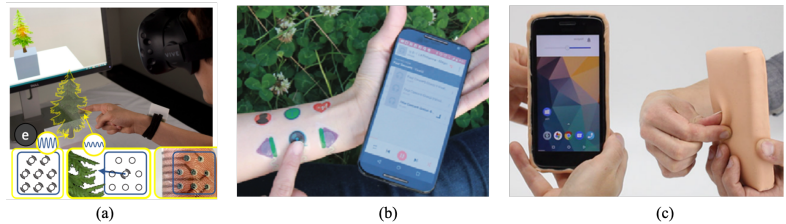

Fig. 8. Prototypes of Epidermal Interfaces. a) an array of a feel-through interface for electro-tactile output on the user's skin allows users to get notification about virtual touch with digital overlaid projected on mobile headsets [85]; b) replaces the direct manipulation on touchscreen interfaces by pressing the self-defined buttons on the user's forearm [72]; c) A skin-On Interface (artificial skin) serve as a new form of input gestures for interface control (e.g., controlling music volume) and emotional communication (e.g., with virtual agents) [97].

### 3.3.2 Epidermal Interfaces.
Humans have a biological predisposition to form attachments with inanimate objects. The interaction on our skin surface (Figure 8) would enhance user engagement [97], with the following prominent features. First, the interactive surface is highly available for swift and subtle interaction on various body location [71]. Second, its thin and conformal form factor [72] results in lightweight, natural, and highly sensitive surfaces. These unique aspects open up various research opportunities in the context of HCI considering the highly mobile scenarios in the urban environment, although this domain significantly relies on material science [128]. Third, due to the wide range of feedback clues, skin-based input owns a higher degree of sense of agency than other input alternatives, including keyboards and voice commands. The sense of agency is defined as the ease of controlling one's own actions and their outcomes [156]. Speech interfaces provide auditory and proprioceptive feedback, while keyboards offer auditory, proprioceptive, visual and haptic feedback. On top of such feedback, skin-based input leverages touch feedback on our bodies [155]. Engineering studies on epidermal devices facilitate the research on skin-based sensing and interaction design. The design space of on-skin interfaces varies, in terms of the pigmentation, texture, thickness of the artificial skins [97], topology [69], natural tactile perception [65], body locations, cosmetic effects [71], artistic factors, and social acceptance [81].

The most recent works demonstrate practical use-cases of epidermal devices for augmenting the human skin surface with the capability of interacting with digital overlays, far more than capacitive buttons for mobile device interaction [94]. Tacttoo [72] serves as a user feel-through interface in AR supported by tactile feedback, apart from a large number of studies on the see-through and hear-through interfaces. An alternative approach of providing tactile feedback is to manipulate the user's hairs on the skin by varying the magnetic fields [98]. Skintillates [85], mincing the centuries-old tattoos, presents embedded LED displays and strain-sensitive sensing units, which enable users to get notifications and perform inputs on artistic skin surfaces. The tattoo with LED can be highly personalized with customized pictures such as dragons, flowers, kites, butterflies, and app icons. The two above examples show that the epidermal devices are getting mature and ready for in-the-wild usage. However, existing epidermal devices share common limitations such as limited lifespan and sustainability [72].

Tactile feedback on the user's fingertips enables swift and responsive interaction with the non-touchable digital overlays in AR. It opens the possibility of dealing with daily objects, including graphical interfaces on papers and walls. The epidermal devices can be extended to daily objects, robotic arms, and drones. In [97], an artificial skin attached to daily objects allows users to perform interface control (e.g., touch, pitch, hard press) and emotional communication to the objects, for instance, slap in anger situations or finger tapping for seeking attention. This facilitates emotional communication by enabling users to show implicit expression to the virtual agent [101], e.g., slapping the artificial skin for communicating the user's anger to the virtual agents. To conclude, interaction design on epidermal interfaces is at the nascent period of development and implementation. Supported by in-depth and well-established studies on the finger interaction techniques [4, 53, 60, 61] and the corresponding interaction design within the finger space [5, 68, 99, 119], easy-to-care and miniature-size epidermal interfaces are good candidate for city-wide mobile interaction, under the premise that increasingly flexible, robust and scalable material options are available [130], and the battery can sustain day-long scenarios. For instance, Tip-Tap [99] demonstrates a battery-free technique for two-finger interaction between the thumb and the index finger, in which contactless and unobtrusive radio waves energize two RFID tags attached on the fingers.

### 3.4 Input Bandwidth

Although hybrid interaction techniques offer attractive incentives such as high mobility as well as intuitive and subtle interaction/interfaces, the key issues of these works are the significantly lower input bandwidth than traditional devices (e.g., the mouse and keyboard duo). We attempt to provide evidence to support this statement through the well-defined problem of text entry that is characterized by a highly consistent evaluation methodology. Text entry involves the repetitive selection of character keys on the keyboard. The performance of this repetitive selection is evaluated in Words-Per-Minute (WPM) [66]. It is important to note that WPM is only an indicator reflecting the speed of the text entry but not the text entry quality (e.g., input accuracy) in various complexity of the textual bodies. However, the WPM indicator can be analogous to the speed of pointing and selecting in the target acquisition problem. In a strict sense, WPM can only respond to the scenarios in which a user highly focuses on repetitive key/button selections, but not serve as a convincing metric for other tasks such as 3D object manipulation and exploring other user affordance in AR. As the interfaces shrink and even disappear, the text entry performance considerably degrades (Figure 9). Normal users with physical keyboards usually achieve 52 WPM, and professional users can hit up to 100 WPM. A study involving 37,000 participants from 160 countries showed that users tap words at an average text entry rate of 38 WPM on smartphones, while the most efficient typist can achieve 85 WPM [136]. The performance gap in text entry rate is closing as the auto-correction feature saves the user's time from fewer strokes on the keyboard.

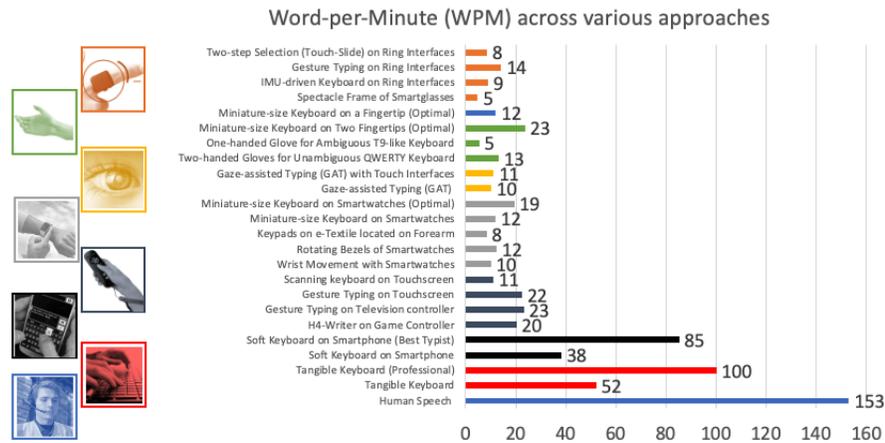

Fig. 9. Input bandwidth across various approaches and interfaces; Eight categories represented by different colours: Natural Speech: Blue; Tangible Keyboard: Red; Soft Keyboard: Black; Hand-held Devices: Dark blue; Interfaces on Forearms: Grey; Gaze: Yellow; Finger Space: Green; Ring-form Devices: Orange.

However, other highly mobile alternative solutions suffer from much larger performance gaps. We illustrate this trend in descending performance and shrinking form-size. First, regarding hand-held devices, users perform thumb interaction on a 4-button layout named H4-Writer on a game controller, achieving almost 20 WPM in the final session [82], while users with a television remote reach an average text entry rate of 23.00 WPM [64]. Alternatively, gesture typing [150] and scanning keyboard [181] on touchscreens can achieve 22.3 WPM and 11.0 WPM, respectively. Second, keyboards on smart wristbands and smartwatches enable users to perform index finger-to-arm interaction on the touchscreen keyboards and makes character inputs on smartglasses [77]. Their modalities vary from wrist movement (9.90 WPM) [66], rotating the bezel areas of the smartwatches (12.25 WPM) [73], e-textile on forearm position (8.11 WPM) [148], as well as tapping on the touchscreens (10.6 [83] – 11.73 [77]). An optimal layout can further push the speed limit to 19.24 WPM [67]. Third, gaze typing is an alternative solution for subtle interaction resulting in a text entry rate of 10.15 WPM [88]. Multi-modal gaze typing systems can significantly improve the rate. GAT is an hybrid system combining touchpad and eye tracking, leading to 11.04 WPM, 25.31% faster than gaze-only typing. Finally, wearable interfaces include two-handed gloves (13.0 WPM) [53], one-handed gloves (4.60 WPM [135] – 4.70 WPM [61]) with additional modality (5.12 WPM [4]), e-skin on fingertips (TipText – 11.90 WPM [5]), the spectacle frame of smartglasses (4.67 WPM [74]) and ring-form devices with various approaches such as IMU-driven selection (8.80 WPM [180]), two-step touch-and-slide selection (8.46 WPM [183]), and word gesture typing (14.00 WPM [182]. The exceptionally high performance on Tiptext [5] is supported by its optimized layout and leverages the highly dexterous index fingers, demonstrating the possibilities of sophisticated engineering and well-designed ergonomics can alleviate the performance gap on the wearable interfaces. Although wearable interfaces display a limited bandwidth, these interfaces are highly capable of handling target acquisition tasks in an accurate and swift manner, e.g., selecting big items and interacting with context-aware data at low interaction cost i.e., within several clicks. Nevertheless, we can leverage speech-based interfaces (up to 153 WPM [184], subject to recognition errors [222]) to fulfil the gaps of high-volume information in specific use cases (e.g., texting a message to your friend(s), talking to a person in a conference call), and other application scenario that are tolerant to speech recognition errors.

## 3.5 Conversational User Interface

The technological advancements in natural language processing, far-field microphone, as well as more diversified user interfaces (e.g., chatbots [109, 116], robots [143], virtual characters [101, 107] like dogs [166] and tutors [167]), allowed CUIs to proliferate [104, 116]. The applications are multitudinous including task management [105] and news delivery [118], micro-task crowd-sourcing [116], video commenting [117], in-vehicle interface controls [123], elderly care, unmanned aerial vehicle (UAV) [153], etc. Among these applications, three types of conversational interfaces frequently appear in the literature [103], which demonstrate a tendency towards multi-modal inputs [108], from voice-based commands [131, 153], interactive dialogues in chatroom alike interfaces [108, 116], to embodied virtual agents [101, 107, 154].

Voice-based commands enable users to naturally interact with the digital contents in AR, for instance, choosing and manipulating 3D graphics with voice commands [114], indicating the longitudinal (forward and backward), vertical (upward and downward), lateral (right and left) movements for flying drones [153], and making appointment in calendar [131]. Although the voice dialogues between the human user and the system pose issues of ineffective information display, hard to review and edit [138] and social acceptance [52], the voice commands are particularly useful for interaction with AR objects and other smart objects like UAV, as users can obtain instant feedback from the target objects. However, voice dialogues are usually commented as black-box systems, and users are unable to make accurate judgements about system capability. As a result, misaligned user expectation occurs [112]. Users feel more uncertain with more complicated and important tasks without appropriate visual confirmations to voice commands [112].

As voice dialogues are transient, the dialogue with systems can be represented as turn-taking GUIs [113], like a chatroom. Visual confirmation serves as a black-and-white proof and enables users to track and trace the dialogues [115], building higher robustness and user experience [108]. Nevertheless, the GUIs are not a silver bullet for better user experience. Indeed, users frequently fall into "guessing" situations [106] due to the communication obstacles of NLP "misheard" [131]. The growing popularity of virtual reality and AR has brought new needs for HCI research on conversational user interfaces since it has bought unexplored patterns of human interactions with computers [101, 107]. Consequently, the CUIs are further evolving to an alternative form of man-like virtual and intelligent agents from 2D GUIs. In the context of HCI, the most recent studies on CUIs have investigated the user perceptions and social perspectives, including the agent's shape and size [154], personality traits [101], playfulness [110], sense of agency (the experience of controlling one's own actions and their outcomes) [155, 156], privacy [111], relationship building with human users [102] trust [139] and empathy [143]. However, the allocation of the screen content on smartglasses are unresolved in CUIs, that is, what conversational contents should be present on the limit-size screen real estate on AR smartglasses? Furthermore, if the virtual and intelligent agents occupy the majority of the screen real estate, how to maintain a reasonable visual confirmation to avoid the pitfalls of transient records appeared in merely voice dialogues? Also, it is crucial to reserve the screen real estate for the primary working tasks in the physical environment. Displaying all the dialogues would hinder the user interaction with the physical world.

As mentioned in Section 3.4, there is a huge performance gap between the traditional interfaces and the miniature-size on-body interfaces. Conversational user interfaces can possibly compensate for the high-volume input bandwidth. The input capabilities should achieve higher coverage, and multi-modal inputs are a promising direction [1]. As for our proposed Human-City Interaction, the low-bandwidth input interfaces should work complementary with a high-bandwidth input channel. Conversational user interfaces (CUI) are thus a good candidate natural interfaces. CUIs significantly occupy the user's cognitive resource governing the short-term memory, working memory, recall, and speech capabilities, and thus deteriorates the problem-solving ability especially for dual-task scenarios, according to the classic model [138]. However, we argue that the availability of miniature-size interface requiring hand-eye coordination can alleviate the user burdens in the low-bandwidth case. CUIs are more adapted for high-bandwidth tasks characterized by a high level of user attention [40].

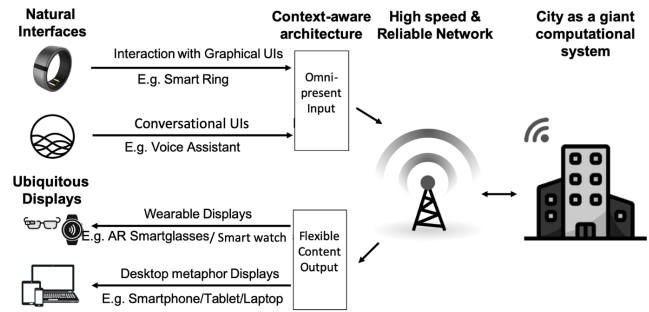

Fig. 10. A Framework of Human-City Interaction from the perspective of HCI; The city-systems offer various applications and no longer limit to the sedentary interactions with desktop computers.

Although we are not yet in a position to offer definitive answers to this question, we see a lot of research opportunities on the design of CUIs driven by the constrained scenarios of mobile AR (e.g., screen real estate [2], user cognitive loads during dual-tasks [14, 144]) for Human-City Interaction with various objects, i.e., tangible and embodied interaction [130] and drones and robots [141]. Also, the design of interaction techniques will soon arrive at the confluence between the input channels of the on-body small-size interfaces, CUIs, and their synergy from multi-modal applications. For example, the augmented sensory capabilities from epidermal and on-body interfaces can augment the conversation design by knowing the environmental context in the surroundings of the users. In WorldGaze [191], the conversational agent leverages the smartphone camera to enhance its context awareness, where the enhanced knowledge of the user's situation makes more intuitive user interaction with the conversational interfaces. As shown in Table 2 (Appendix), the design space of CUIs should consider the wider aspects of networking and AR interfaces. Finally, more attention should be paid to the gap between the laboratory setting and the real-life usage [110].

## 4  RESEARCH AGENDA: AUGMENTED REALITY HUMAN-CITY INTERACTION

Throughout our discussion in Sections 2 and 3, we saw that the HCI research community is eager to design AR interfaces and interaction techniques towards minimal and mobile wearable devices, establishing their essentiality in the era of Human-City interaction. On minimal interfaces such as finger-addendum devices, the usual operations on touchscreen interfaces – including tap, pan, swipe, as well as two-finger zoom – can be substituted by the user's body movements and subtle gestures. With this knowledge, we have to employ alternative input modalities (e.g., IMU-induced wrist movements and force-based finger presses) other than solely touch-based interaction [199], in case some operations require touch interaction areas larger than the usual size of finger-addendum devices. Smartglasses users in city-wide interaction scenarios require interaction techniques supporting a high level of mobility. Thus, the existing works on miniature-size interfaces demonstrate the possibility of highly mobile yet concealed operations on the user's bodies and daily objects. Such a small area has a limited capability of fulfilling the above usual interaction techniques. The collected studies provides us sound justifications for matching various operations with highly mobile and miniature devices. Nevertheless, a systematic evaluation of such interaction areas is still missing. It is crucial to systematically quantify the requirements of interaction areas with the interaction techniques [60].

We see that natural interfaces (Figure 10) leveraging the theories of embodiment (Section 3.2) can achieve a reasonable input bandwidth on AR headsets. Miniature interfaces allow the users to precisely point and select the targets in AR, but their prolonged and repetitive use rapidly leads to user fatigue [1, 60]. Moreover, as mentioned in Section 3.4, we foresee that such interfaces still present a significant performance gap with other more traditional interfaces such as touchscreens and tangible keyboards. We thus require alternative interaction methods for high-volume inputs, and conversational user interfaces (CUIs) can become a good candidate to work complementary with such interaction techniques of limited bandwidth. However, we identify improvement areas for more natural CUIs. First, the conversational agent's response conversation cannot be interrupted even though the user is not getting a desired answer in the conversation [114]. Second, the agents only respond to the current questions from a user, which means the prior questions cannot be considered as a continual conversation [108]. Third, CUIs are less transparent than GUIs. CUIs act as black-box-like agents, and the users cannot explicitly know the functions and their maximum extents [105]. The agent can become more natural in the above three areas by leveraging the city smartness and the corresponding context in the conversation.

On the other hand, the display of AR content on AR headsets, one of the many ubiquitous displays (Figure 10) of city-systems, poses multiple challenges. First, current AR displays suffer from a limited field of view, and thus a limited screen real-estate [2]. Besides, using such hardware can generate a significant visual [7] and cognitive load [8]. Rather than listing comprehensive yet less-than-necessary information as it is the case on desktop computers, user interaction should leverage the smart city's context awareness to organize the information extraction and to present simplified interfaces depending on the user context [6]. As AR content is meant to integrate the physical world, users are often involved in some physical tasks augmented with AR contents. Therefore, the visual [7] and cognitive loads [8] on the smartglasses, with the likelihood of social interruptions in the user's surrounding, should be considered in the AR interface design.

On the basis of the aforementioned motivations, Figure 10 sketches out the complementary roles of miniature interfaces and CUIs. Such technologies can be brought together efficiently by a context-aware framework leveraging the city-system for user-centric interaction. We have to explore the mutual design space of miniature interfaces and CUIs under the backbone of such a context-aware framework to derive insightful and practical design implications for Human-City Interaction through hardware-constrained wearable computers. This context-aware architecture can further drive multi-modal interaction in which the shift from one modality to another can be designed by intelligent decision with a high level of context-awareness. In addition, the AR user interaction will be employed in public space, and thus speech-based interfaces should own the capability of handling privacy leakage issues (e.g., eavesdropping by bystanders). A context-aware framework should offer an adaptive switch between interaction modals in some privacy-sensitive environments. For instance, the interaction method will switch from conversational to graphical UIs during authentication events in public space but not in a private room (e.g., alone at home).

We formulate the above conjecture of multi-modal inputs, and construct our arguments for Human-City Interaction in the perspective of AR user interaction design, according to the most up-to-date works. We consider the city as a computational and intelligent system where the different components are connected with reliable high-speed networks. As indicated by Figure 10, the computationally-demanding tasks can be offloaded to cloud and edge servers adjacent to the users, while wearable computers operate as interaction tokens between users and city-systems. The interaction tokens are composed of natural interfaces under the combination of speech-based commands plus a miniature addendum (e.g., a ring) that offers an extra area to exert the body-centric interaction techniques (e.g., on-body interaction / hand gestural inputs). The futuristic scenes (Figure 11) further illustrates the projected impacts of employing ubiquitous user interaction with the city systems in various service points (e.g., supermarkets, commercial streets, and catholic churches) through seamless and lightweight interaction tokens supported by natural interfaces. In the next section, we bring the augmented reality interfaces, user interaction techniques and the networking issues together for the city-wide user interaction.

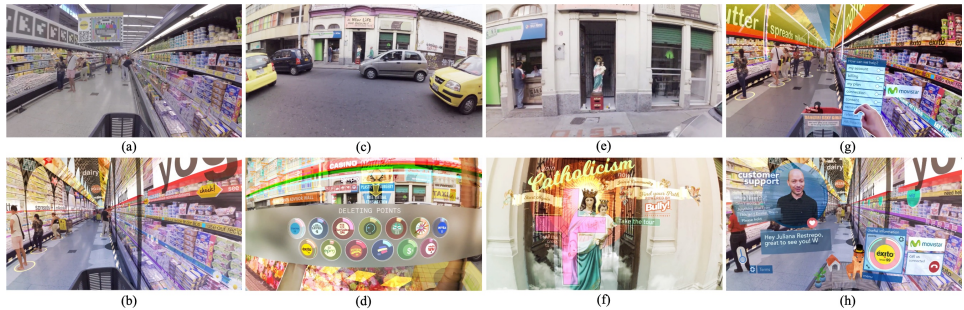

Fig. 11. Several projected scenes when AR becomes the omnipresent media (i.e., user interfaces) in our urban environments. a) an example of indoor scenario, where supermarket with QR codes (i.e., AR marker); b) an augmented scene of AR-enabled supermarket on the physical surrounding of (a); c) an outdoor scenario – a street; d) an augmented view of the street in (c), relying on geographical locations and user contexts to decide what services to be offered; e) an inanimate object of a catholic statue; f) converting (e) into lively media regarding a religious organisation; g) Hand gestural interactions on an AR menu (accurate pointing and selecting items, as an example of natural interfaces); and h) conversational interfaces for enquiring the account status in the supermarket (high-volume inputs); Image sourced from a short movie "*Hyper-Reality*", created by Keiichi Matsuda (http://km.cx/projects/hyper-reality).

## 4.1 AR, Interaction and Networks

AR for Human-City Interaction relies on large scale deployment. Applications are not anymore limited to a single place, but cover the entire city's landscape. As such, we consider the smart city as a single massive system. The various components of this system are connected through wireless networks instead of high-speed buses. The network capability of the system may thus become one of the major bottlenecks. Depending on the internal structure of the city-system, low network bandwidth or high latency at a single point can considerably degrade the entire system performance, similarly to how a low-speed hard drive can considerably reduce the perceived operating speed of a desktop computer.

Regarding interaction, even in basic applications, AR often relies on heavy computations that lightweight wearables such as smartglasses can barely handle [163]. Image recognition, for instance, requires large databases of feature points or resource-intensive models that cannot be realistically embedded in constrained headsets [189]. Scaling up to thousands of square kilometres areas will only increase the issue. In the case of smart cities, the amount of data to interact with is so large that large-scale context-awareness is required for intelligently selecting the data to display to the user [190]. For instance, connected vehicles may share vision to improve the awareness of the driver [203, 204]. In such scenario, every unnecessary information will distract the driver instead of improving his or her awareness of danger. Offloading computations to remote machines is the generally accepted response to these problems [10]. When the AR smartglasses maintain a lightweight interface, user interaction across the network with the city-system will encounter unavoidable latency. The interaction techniques, as basic as pointing and selecting objects, need to address the issues [158]. Latency in AR is a critical parameter, and alignment issues between the virtual content and the physical world arise for motion-to-photon latencies as low as 20 ms [11]. Besides, users have variable tolerance to drops in performance depending on the application. For instance, when streaming videos, users tend to prefer a fluid, yet degraded video quality to a choppy high-quality stream. Immersive interfaces in AR are a particular case of AR applications as they rely heavily on detecting markers, objects, and context clues from the video feed of the camera to fix their position. These interfaces require timely interaction for seamless user experience, and might degrade quickly with the delay between user input and the corresponding feedback. A typical example of this phenomenon is text input, where the user performs a quick succession of repetitive actions over a prolonged period. Even slight variations of the motion-to-photon latency may destabilize the user and lead to decreased performance with higher error rates. Finally, in our daily lives, the natural conversation creates the necessity for context-awareness, where users have habits of mapping 'this' and 'that' to certain objects. Intelligent interfaces also require context-awareness for intuitive operation. However, a subtle change in context along with network latency may distort the user meaning.

Designing such interfaces requires to strike a balance between computation times, network and computation latency, and Quality of Experience. In other words, we need to ask *which parameters can we afford to degrade when a sudden change in the network conditions happens without affecting the user's Quality of Experience.* We realize that there is no one-size-fits-all answer to this question. However, it is possible to draw some general recommendations. In general, it is preferable to lower the quality (polygons, details, texture quality) of the displayed objects, rather than enabling large motion-to-photon latencies. Similarly, larger latencies in a fluid experience tend to impact the user less than a highly variable or choppy display of the virtual elements. In case it is not acceptable to degrade the user's Quality of Experience, for instance, in safety-related applications, exploiting multiple links and servers can significantly increase the available resources [202]. Besides these general guidelines, the design of immersive AR interfaces should follow a holistic approach, in which every parameter is tailored to the exact requirements and use cases of the interface.

## 4.2 Citizens with Special Needs and Inclusion in Augmented City Design

Apart from the technological challenges between AR and smart cities, AR user interaction design should consider the full spectrum of stakeholders. It should thus not neglect citizens with special needs, including both mental and physical impairments. The application of remote assistance on Google Glass has long been considered a practical and successful application since its launch in 2014 and recently merged with Google Hangout Meets for online communication and work collaboration amongst distal workers in the Covid-19 crisis [206]. When our cities are equipped with increasing sensing ability, the orientations of remote support and user collaborations on AR headsets can become a foundation for designing immersive urban with an additional incentive of promoting empathy among different stakeholders. *Empathy* is *the ability to put oneself in the other's situations, both physically, emotionally, and intellectually* [209]. The tactile paving and braille signage for sightless individuals, as well as stairlifts and ramps for wheelchair patrons in our cities, are traditional facilities of building an inclusive society with individuals with special needs. However, immersive assets as virtual facilities have undefined roles of providing empathy and accessibility functions to individuals with special needs and the aging populations. Further research is needed to empathetically elicit the user needs in immersive urban environments and accordingly provide user interaction design with them, especially from the perspective of assistive technology.

We have seen several relevant examples of leveraging immersive technologies across the reality-virtuality continuum, as follows. Virtual reality has been used to raise the empathy level of collaborators (remote helpers) through sharing a first-person view of an in-situ worker through the worker's headset [208]. Simultaneously, the system is connected to a bundle of sensors allowing the remote helpers to enhance the awareness of the in-situ worker's physiological cues such as facial expression and heartbeat [208]. Also, the participants with virtual reality (VR) headsets throughout serious games can gain experiential understanding of the hardship of the minority and cultivate enduring positive attitudes toward those minority groups. *Becoming Homeless: A Human experience*, developed by Stanford University's Virtual Human Interaction Lab [207], is a VR serious game that allows the public to imagine oneself to spend several days in the life of becoming homeless. Similar

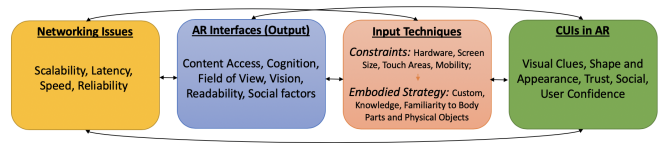

Fig. 12. The four key areas discussed and their high-level issues, acting as a 'think-aloud' tool considering multiple issues of different areas for new research problems.

VR serious games such as *I Am A Man* [218] and *1,000 Cut Journey* [219] offer immersive experiences of the struggles amongst the marginalized people of color in order to raise awareness on the issues of race and racism. Another VR serious game focuses on cultivating the empathy of informal caregivers who look after individuals with dementia. The educational scenarios in VR could strengthen the informal caregivers' understanding of the difficulties of individuals with dementia, such as memory loss and unable to recognize daily objects [253].

Similarly, the immersive experience can deliver a body ownership illusion that allows users to experience how others view and interact with their environment [211]. This concept can be further extended to the group with special needs, allowing the majority (i.e., non-disabled adults) to understand and reflect the design negligence that existed for a long time in our cities. Such negligence include disability (e.g., visually impaired individuals [210]), people of varying heights (e.g., giant and dwarf [213, 214]), and arm lengths [217], discounted physical ability (e.g., kids [215]), and persons with reduced dexterity or dyslexia [216]. Next, the delivery channels of such body ownership illusion can be either purely immersive or hybrid with physical gadgets. For example, the headset user in a visual impairment simulator can see the world with various low vision symptoms such as loss of visual field/ central vision/ visual acuity, distortion, patched vision, and photophobia [210]. Instead of employing a purely immersive experience, physical augmentation also sheds light on the design of immersive environments with empathy. HandMorph [215] is a passive exoskeleton that transforms an adult user's hand to a smaller hand of reduced grasping range. The exoskeleton can effectively communicate with adult designers how a smaller person (e.g., kids) sees the everyday things designed for adults. Accordingly, the designers gain confidence in designing our daily objects through understanding users with different sizes. More examples of physiological, perceptual, and mobility restrictions are documented in [216].

As shown in the examples mentioned above, the virtual contents or immersive experience can be inserted into the physical world as a part of the AR cues. Simultaneously the individuals will receive reminders about the minority groups and people with special needs in context-aware manners [212], i.e., when they meet the person with special needs or give cues to make empathetic responses in the conversations [220]. Moreover, the experience of knowing the difficulties, hurdles, and pain points of others allow us to build the city-system more inclusively. It is important to note that participatory design should move towards the attitude of '*being with*' from '*being like*' to avoid the empathy pitfalls [221]. Finally, the nascent AR is seeking its position and role in our daily tasks and way of life. As various AR contents representing the services of our giant city systems that serve everybody in the social group inside the city, perhaps we can augment our world with empathy and hence secure a niche position of AR for a projected empathy and unite society. Extended Reality (XR) refers to the full spectrum of the reality-virtuality continuum, including AR, VR, and MR. We foresee that AR and empathy in the context of individuals with special needs can potentially converge in a new class under XR, which we dub *Empathetic Reality* (ER).

### 4.3 Research Directions and Opportunities

This article illustrates the conceptual framework of Human-City Interaction primarily from the angles of interfaces and interaction approaches in the smart urban environment. Figure 12 generalizes the issues of the user interaction design for city-wide augmented reality. Accordingly, we outline research opportunities by concatenating the four discussed aspects of networking, outputs, inputs, and CUIs, as shown in Table 2 (Appendix). Finally, we depict the research directions of human-city interaction as follows.

*Mixing Augmented Reality Interfaces with Brick-and-mortar Urban Entities.* We have investigated various concepts and frameworks under the domain of Augmented Reality. We believe that augmented reality (AR) with a high degree of context-awareness could serve as a window bridging smart cities to human users. Instead of constructing 2D UIs based on the traditional WIMP paradigm, the workbench in AR is decentralized into several AR overlays that are closely relevant to the contexts of the physical world. For example, by binding the digital interfaces with physical objects together, Ubii achieves a nearby embodiment to afford user awareness to interact with domain interfaces and objects. The example sheds light on designing digital contents with the focus on the reality-based interaction [57] as well as the user interaction with building infrastructure [195] in our urban environments.

*Towards Scalable AR Interfaces.* Nowadays, we have created massive amounts of data on the web. Leveraging the web data can make AR interfaces more relevant and meaningful to the in-use context, i.e., the user's situations. We can also transform the web platform into easy-to-use AR interfaces with high scalability for the smart city's mass applications. Mobile-to-AR (M2A) [6] works on user-centric AR web browsing and derives the responsive design paradigm from mobile web development to the AR context. Under the premise of minimal interfaces, simplified yet context-aware interfaces reduce unnecessary options, thus improving the user attention and cognitive loads. M2A exploits the visual context to display more user-interested content and enables users to locate relevant information intuitively. Also, developers with M2A need minimal effort to modify the existing website into context-aware AR websites.

After managing the AR contents in the user's interests (i.e., selecting the right content), the next important issue is to put the contents at the right place (i.e., place the right content right) of the citizen behaviors. Moreover, it is worthwhile to mention that user behaviors in our cities are changing. It will be almost infeasible and highly costly if we manually and constantly update the AR cues according to the changing user behavior for every corner of the city. Our smart urban environments are necessary to adapt themselves to the user needs and to provide customized interfaces [201]. The AR cues on smartglasses can become obsolete if they are separated from the user behavior. Thus, presenting the AR cues in such dynamic urban environments in highly scalable manners remains an open question.

*Body-centric Interaction for City-wide Augmented Reality.* Inspired by science fiction movies like **Minority Report**, some existing hand gestural input systems, such as TiPoint [3] and HIBEY [2], enable users to interact with digital overlays through hardware-constrained smartglasses naturally. However, we see the deficiencies in the sole hand gestural system, which is more coarse and less responsive than touch-based interfaces [1]. Nevertheless, we intend to reserve the characteristics of naturalness and intuitiveness with body-centric interaction. We have a strong belief that our body is the readily available resource to compensate for the resource-constrained AR smartglasses. Moreover, we wonder about the possibility of employing hybrid approaches of body-centric and touch-based interaction.

Another more recent work [4] explores the design space of thumb-to-finger space interaction in mobile scenarios and employs sensors augmenting the finger space with touch-sensitive and force-assisted capability. Users with such a thumb-to-finger solution can still apply thumb-based press within the finger space naturally and achieve swift responses from the sensors. We foresee that on-body sensors leveraging natural body movements can become promising solutions for interacting with city-systems with high mobility, albeit the gap of input bandwidth (Figure 9) exists. Additionally, apart from body-centric user interaction with AR digital overlays, the body-centric approaches can be further extended to the user interaction with daily objects [186] and IoT objects [199]. One of such approaches' prominent features is that the users can get rid of carrying numerous controllers and simultaneously achieve convenient and on-demand user interaction with multiple daily and IoT objects.

*Towards Computational and Iterative HCI Studies.* In many of the most recent works, HCI researchers have been interested in dealing with certain interaction design problems computationally [161, 200]. As AR smartglasses pose constrained input and output bandwidths, this is particularly crucial to justify the interfaces and layout design quantitatively. Rather than an opportunistic approach to putting sensors together, the research community figures out these problems incrementally in the evolving loop of user feedback and computational models. Before reaching the final solution, the design space has been investigated in several possibilities after consulting the user feedback in preliminary tests. HIBEY [2] shows the importance of understanding the users' position displacement when picking characters in the holographic environment, and afterward include these erroneous positions in a probabilistic model supporting the keyboard-less and invisible text entry. Additionally, we see a similar exploration process again within a glove project [4]. Before computing the keyboard layouts within the finger space, their pilot study measures the user performance with force-assisted interaction. In on-finger addendum devices for subtle interaction with AR smartglasses [60] and IoT objects [199], a large amount of interaction footprint were collected before reaching the final solutions of AR user interaction techniques. Nevertheless, the existing approaches of computational and iterative studies need a high level of involvement from researchers. The limited human resource of researchers hardly fulfills the growing demands of city-scale experiments. But very few works consider designing urban interactive technologies driven by the proactive user's probes, under the premise that no researcher is directly involved in the design process [205].

## 5 CONCLUSION

With the proliferation of commercial AR headsets, interfaces and interaction techniques will look radically different in the upcoming years. Our immersive urban's future is going to be more interactive, more alive and more multimedia. However, there are still many challenges to be overcome before head-worn AR technology become integrated into our everyday life. We call for a context-aware interaction paradigm supporting multi-modal interfaces at a miniature size, accompanied with conversational user interfaces to support such interactive systems. By surveying the most recent works across AR output (Table 1, Appendix) and input (Table 3, Appendix), we hope to have provided a wider discussion within the HCI community. Through reflecting on the key topics we discussed, we identify the fundamental challenges and research problems (Table 2, Appendix) to shape the future of AR Interaction at the city-wide scale (Human-City Interaction).

## ACKNOWLEDGEMENT

This research has been supported in part by 5G-VIIMA and REBOOT Finland IoT Factory projects funded by Business Finland, the 6G Flagship project, 5GEAR funded by the Academy of Finland (Decision No. 318927), and project 16214817 from the Research Grants Council of Hong Kong.

## REFERENCES

[1] L. H. Lee and P. Hui. Interaction methods for smart glasses: A survey. IEEE Access, 6:28712–28732, 2018.

[2] L. H. Lee, K. Y. Lam, Y. P. Yau, T. Braud, and P. Hui. HIBEY: Hide the keyboard in augmented reality. In 2019 IEEE International Conference on Pervasive Computing and Communications(PerCom, pages 1–10, March 2019.

[3] L. H. Lee, T. Braud, F. H. Bijarbooneh, and P. Hui. Tipoint: Detecting fingertip for mid-air interaction on computational resource constrained smartglasses. In Proceedings of the 23rd International Symposium on Wearable Computers, ISWC '19, pages 118–122 NY, USA, 2019. ACM.

[4] L. H. Lee, K. Y. Lam, T. Li, T. Braud, X. Su, and P. Hui. Quadmetric optimized thumb-to-finger interaction for force-assisted one-handed text entry on mobile headsets. Proceedings ACM Interact. Mob. Wearable Ubiquitous Technol., 3(3):94:1–94:27, September 2019.

[5] Z. Xu, P. Chung Wong, J. Gong, T. Wu, A. S. Nittala, X. Bi, J. Steimle, H. Fu, K. Zhu, and X. Yang. Tiptext: eyes-free text entry on a fingertip keyboard. In Proc of the 32nd Annual ACM Symp. on User Interface Soft. and Tech., UIST '19, pages 883–899 NY, USA,2019. ACM.

[6] K. Y. Lam, L. H. Lee, T. Braud, and P. Hui. M2a: A framework for visualizing information from mobile web to mobile augmented reality. In 2019 IEEE International Conference on Pervasive Computing and Communications (PerCom, pages 1–10, March 2019.

[7] I. Chaturvedi, F. H. Bijarbooneh, T. Braud, and P. Hui. Peripheral vision: A new killer app for smart glasses. In Proceedings of the 24th International Conference on Intelligent User Interfaces, IUI '19, pages 625–636 NY, USA, 2019. ACM.

[8] D. Lindlbauer, A. M. Feit, and O. Hilliges. Context-aware online adaptation of mixed reality interfaces. In Proc of the 32nd Annual ACM Symp. on User Interface Soft. and Tech., UIST '19, pages 147–160 NY, USA, 2019. ACM.

[9] Z. Huang, W. Li, and P. Hui. Ubii: Towards seamless interaction between digital and physical worlds. In Proc ofthe 23rd ACM Inter. Conf. on Multimedia, MM '15, pages341–350 NY, USA, 2015. ACM.

[10] T. Braud, F. H. Bijarbooneh, D. Chatzopoulos, and P. Hui. Future networking challenges: The case of mobile augmented reality. In 2017 IEEE 37th International Conference on Distributed Computing Systems (ICDCS), pages 1796–1807, June 2017

[11] T. Braud, T. Kämäräinen, M Siekkinen and P. Hui. Multi-Carrier Measurement Study of Mobile Network Latency: The Tale of Hong Kong and Helsinki. In 2019 The 15th International Conference on Mobile Ad-hoc and Sensor Networks, Shenzhen, China.

[12] Y. Matsuura, T. Terada, T. Aoki, S. Sonoda, N. Isoyama, and M. Tsukamoto. 2019. Readability and legibility of fonts considering shakiness of head mounted displays. In Proc of the 23rd Inter. Symp. on Wearable Computers (ISWC '19). ACM NY, USA, 150–159.

[13] M. Nakao, T. Terada, and M. Tsukamoto. 2014. An Information Presentation Method for Head Mounted Display Considering Surrounding Environments. In Proc of the 5th Augmented Human Inter. Conf. (AH '14). ACM NY, USA.

[14] K. Tanaka, Y. Kishino, M. Miyamae, T. Terada, and S. Nishio. 2007. An Information Layout Method for an Optical See-Through HMD Considering the Background. In Proc of the 11th IEEE Inter. Symp. on Wearable Computers (ISWC '07). IEEE NY, USA.

[15] M. Zhao, H. Qu, and M. Sedlmair. 2019. Neighborhood Perception in Bar Charts. CHI '19. ACM NY, USA, Paper 232, 1–12.

[16] Y. Lin, L. Hsu, and M. Y. Chen. 2018. PeriTextAR: utilizing peripheral vision for reading text on augmented reality smart glasses. In Proc of the 24th ACM Symp. on Virtual Reality Soft. and Tech. (VRST '18). ACM NY, USA, Art. 110, 1–2.

[17] J. Franco and D. Cabral. 2019. Augmented object selection through smart glasses. (MUM'2019).

[18] E. M. Klose, N. A. Mack, J. Hegenberg and L. Schmidt. 2019. Text Presentation for Augmented Reality Applications in Dual-Task Situations. 2019 IEEE Conf. on Virtual Reality and 3D User Interfaces (VR) (VRST'19): 636-644.

[19] X. Zhao, K. Go, K. Kashiwagi, M. Toyoura, X. Mao and I. Fujishiro. 2019. Computational Alleviation of Homonymous Visual Field Defect with OST-HMD: The Effect of Size and Position of Overlaid Overview Window. 2019 Inter. Conf. on Cyberworlds (CW) (2019): 175-182.

[20] R. Rzayev, S. Mayer, C. Krauter and N. Henze. 2019. Notification in VR: The Effect of Notification Placement, Task and Environment. CHI PLAY '19.

[21] R. Rzayev, P. W. Wozniak, T. Dingler and N. Henze. 2018. Reading on Smart Glasses: The Effect of Text Position, Presentation Type and Walking. CHI '18.

[22] W. S. Lages, and D. A. Bowman. 2019. Walking with adaptive augmented reality workspaces: design and usage patterns. IUI '19 .

[23] D. Lindlbauer, A. M. Feit and O. Hilliges. 2019. Context-Aware Online Adaptation of Mixed Reality Interfaces. UIST '19.

[24] D. Boyarski, C. Neuwirth, J. Forlizzi, and S. H. Regli. 1998. A Study of Fonts Designed for Screen Display. CHI '98. ACM Press/Addison-Wesley Publishing Co. NY, USA, 87–94.

[25] W. Xu, H. Liang, A. He and Z. Wang. 2019. Pointing and Selection Methods for Text Entry in Augmented Reality Head Mounted Displays. 2019 IEEE Inter. Symp. on Mixed and Augmented Reality (ISMAR), Beijing, China, 2019, pp. 279-288.

[26] K. Arthur. 1996. Effects of field of view on task performance with head-mounted displays. CHI '96.

[27] S. H. Chua, S. T. Perrault, D. J. C. Matthies and S. Zhao. 2016. Positioning Glass: Investigating Display Positions of Monocular Optical See-Through Head-Mounted Display. ChineseCHI 2016.

[28] K. Tanaka, Y. Kishino, M. Miyamae, T. Terada and S. Nishio. 2008. An information layout method for an optical see-through head mounted display focusing on the viewability. 2008 7th IEEE/ACM Inter. Symp. on Mixed and Augmented Reality, Cambridge, pp. 139-142.

[29] T. Ianchulev, D. S. Minckler, H. Hoskins, et al. 2014. Wearable technology with head-mounted displays and visual function. JAMA 312, 17 (2014), 1799–1801.

[30] B. Moggridge. Designing Interactions. The MIT Press, 2006

[31] Here's why your computer has a mouse, according to Steve Jobs in 1985. https://www.cnbc.com/2018/05/21/why-your-computer-has-a-mouse-according-to-steve-jobs.html, 2018.

[32] Smartphone User Interface. https://www.textrequest.com/blog/history-evolution-smartphone/, 2013

[33] The History and Evolution of the Smartphone: 1992-2018. https://www.textrequest.com/blog/history-evolution-smartphone/, 2012

[34] The Touching History of Touchscreen Tech. https://mashable.com/2012/11/09/touchscreen-history/K.XAjA.Rsqg, 2012

[35] S. Rothe, K. Tran, and H. Hussmann. 2018. Dynamic Subtitles in Cinematic Virtual Reality. In Proc of the 2018 ACM Inter. Conf. on Interactive Experiences for TV and Online Video (TVX '18). ACM NY, USA, 209–214.

[36] A. Woodham, M. Billinghurst, and W. S. Helton. 2016. Climbing With a Head-Mounted Display: Dual-Task Costs," (eng), Human factors, vol. 58, no. 3, pp. 452–461.

[37] S. Fadden, C. D. Wickens, and P. Ververs. 2000. Costs and Benefits of Head up Displays: An Attention Perspective and a Meta Analysis. In Proc. of the 2000 World Aviation Conference, San Diego, CA.

[38] J. Orlosky, K. Kiyokawa, and H. Takemura. 2014. Managing mobile text in head mounted displays. SIGMOBILE Mob. Comput. Commun. Rev., vol. 18, no. 2, pp. 20–31.

[39] E. Fischer, R. F. Haines, and T. A. Price. 1980. Cognitive Issues in Head-Up Displays. NASA, Technical Report 1711.

[40] K. W. Gish and L. Staplin. 1995. Human Factors Aspects of Using Head Up Displays in Automobiles: A Review of the Literature. Report No. DOT HS 808 320. Washington, DC: U.S. Department of Transportation - National Highway Traffic Safety Administration.

[41] S. Hudson, J. Fogarty, C. Atkeson, D. Avrahami, J. Forlizzi, S. Kiesler, J. Lee, and J. Yang. 2003. Predicting Human Interruptibility with Sensors: A Wizard of Oz Feasibility Study. CHI '03. ACM NY, USA, 257–264.

[42] M. A. McDaniel, G. O. Einstein, T. Graham, and E. Rall. 2004. Delaying execution of intentions: Overcoming the costs of interruptions. Applied Cognitive Psychology 18, 5 (2004), 533–547.

[43] M. Muhlenbrock, O. Brdiczka, D. Snowdon, and J-L Meunier. 2004. Learning to detect user activity and availability from a variety of sensor data. In Pervasive Computing and Communications, 2004. PerCom 2004. Proc of the Second IEEE Annual Conf. on. IEEE, 13–22.

[44] D. Newtson. 1973. Attribution and the unit of perception of ongoing behavior. J. Personality and Social Psychology 28, 1 (1973), 28.

[45] T. Tanaka and k. Fujita. 2011. Study of User Interruptibility Estimation Based on Focused Application Switching. In Proc of the ACM 2011 Conf. on Computer Supported Cooperative Work (CSCW '11). ACM NY, USA, 721–724.

[46] S. T. Iqbal and B. P. Bailey. 2005. Investigating the Effectiveness of Mental Workload As a Predictor of Opportune Moments for Interruption. In CHI '05 Extended Abstracts on Human Factors in Computing

[47] C. George, P. Janssen, D. Heuss, and F. Alt. 2019. Should I Interrupt or Not? Understanding Interruptions in Head-Mounted Display Settings. In Proc of the 2019 on Designing Interactive Systems Conference (DIS '19). ACM NY, USA, 497–510.

[48] J. M. Hudson, J. Christensen, W. A. Kellogg, and Thomas Erickson. 2002. "I'D Be Overwhelmed, but It's Just One More Thing to Do": Availability and Interruption in Research Management. CHI '02. ACM NY, USA, 97–104.

[49] M. Gattullo, A. Uva, M. Fiorentino, J. Gabbard. (2015). Legibility in Industrial AR: Text Style, Color Coding, and Illuminance. Computer Graphics and Applications, IEEE. 35. 52-61.

[50] M. Khademi, H. Mousavi Hondori, C. Lopes. (2012). Optical Illusion in Augmented Reality. Proc of the ACM Symp. on Virtual Reality Soft. and Tech., VRST.

[51] 1 GHz: Intel claims it was first. https://www.zdnet.com/article/1-ghz-intel-claims-it-wasfirst/,2000.

[52] Y. T. Hsieh, A. Jylhä, V. Orso, L. Gamberini, and G. Jacucci. Designing a willing-to-use-in-public hand gestural interaction technique for smartglasses. In Proc of the 2016 CHI Conf. on Human Factors in Comp. Sys., CHI '16, pages 4203–4215, San Jose, California, USA, 2016. ACM.

[53] E. Whitmire, M. Jain, D. Jain, G. Nelson, R. Karkar, S. Patel, and M. Goel. Digitouch: Reconfigurable thumb-to-finger input and text entry on head-mounted displays. volume 1, pages 113:1–113:21, Maui, Hawaii, USA, September 2017. ACM.

[54] D. O'Sullivan and T. Igoe. Physical Computing: Sensing and Controlling the Physical World with Computers. Course Technology Press, Boston, MA, United States, 2004.

[55] S. Leigh, H. Sareen, H. Kao, X. Liu, and P. Maes. Body-borne computers as extensions of self. Computers, 6(1), 2017.

[56] K. J. Friston, J. Daunizeau, J. Kilner, and S. J. Kiebel. Action and behavior: a free-energy formulation. Biological Cybernetics, 102(3):227–260, Mar 2010.

[57] R. J.K. Jacob, A. Girouard, L. M. Hirshfield, M. S. Horn, O. Shaer, E. T. Solovey, and J. Zigelbaum. Reality-based interaction: A framework for post-wimp interfaces. In Proc of the SIGCHI Conf. on Human Factors in Comp. Sys., CHI '08, pages 201–210, Florence, Italy, 2008. ACM.

[58] A. B. Garg. Embodied cognition, human computer interaction, and application areas. In T.H. KIM, et al. Computer Applications for Web, Human Computer Interaction, Signal and Image Processing, and Pattern Recognition, pages 369–374, Berlin, Heidelberg, 2012. Springer Berlin Heidelberg.

[59] P. Dourish. AI Being-in-the-World: Embodied Interaction. MITP, 2004.

[60] L. H. Lee, Y. Zhu, Y. P. Yau, T. Braud, X. Su and P. Hui. 2020. One-thumb Text Acquisition on Force-assisted Miniature Interfaces for Mobile Headsets. In Proc. IEEE Inter. Conf. on Pervasive Computing and Communications (PerCom 2020)

[61] P. C. Wong, K. Zhu, and H. Fu. 2018. Fingert9: Leveraging thumb-to-finger interaction for same-side-hand text entry on smartwatches. In Proc of the 2018 CHI Conf. on Human Factors in Computing System

[62] K. Johnson (2019). "Yann LeCun: AR glasses will be the killer app of energy-efficient machine learning," VentureBeat. https://venturebeat.com/2019/12/17/yann-lecun-ar-glasses-will-be-the-killer-app-of-energy-efficient-machine-learning/ (Access: 9 Jan 2020).

[63] A. Sharma, J. S. Roo and J. Steimle. 2019. Grasping Microgestures: Eliciting Single-hand Microgestures for Handheld Objects. CHI '19, Paper 402, 13 pages.

[64] S. Zhu, J. Zheng, S. Zhai and X. Bi. 2019. i'sFree: Eyes-Free Gesture Typing via a Touch-Enabled Remote Control. CHI '19, Paper 448, 12 pages.

[65] A. S. Nittala, K. Kruttwig, J. Lee, R. Bennewitz, E. Arzt, J. Steimle. 2019. Like a Second Skin: Understanding How Epidermal Devices Affect Human Tactile Perception. CHI '19, Paper 380, 16 pages.

[66] J. Gong, Z. Xu, Q. Guo, T. Seyed, X. Chen, X. Bi and X. Yang. 2018. WrisText: One-handed Text Entry on Smartwatch using Wrist Gestures. CHI '18, 1-14.

[67] R. Qin, S. Zhu, Y. H. Lin, Y. J. Ko and X. Bi. 2018. Optimal-T9: An Optimized T9-like Keyboard for Small Touchscreen Devices. In Proc of the 2018 ACM Inter. Conf. on Interactive Surfaces and Spaces (ISS '18), 137-146.

[68] M. Soliman, F. Mueller, L. Hegemann, J. S. Roo, C. Theobalt, and J. Steimle. 2018. FingerInput: Capturing Expressive Single-Hand Thumb-to-Finger Microgestures. In Proc of the 2018 ACM Inter. Conf. on Interactive Surfaces and Spaces (ISS '18), 177-187.

[69] Y. Wang, S. Luo, H. Gong, F. Xu, R. Chen, S Liu and P. Hansen. 2018. SKIN+: Fabricating Soft Fluidic User Interfaces for Enhancing On-Skin Experiences and Interactions. CHI EA '18, 1-6.

[70] M. Weigel and J. Steimle. 2017. DeformWear: Deformation Input on Tiny Wearable Devices. Proc. ACM Interact. Mob. Wearable Ubiquitous Technol. 1, 2, Art. 28 (June 2017), 23 pages.

[71] M. Weigel, T. Lu, G. Bailly, A. Oulasvirta, C. Majidi and J. Steimle. 2015. iSkin: Flexible, Stretchable and Visually Customizable On-Body Touch Sensors for Mobile Computing. CHI '15, 2991-3000.

[72] A. Withana, D. Groeger and J. Steimle. 2018. Tacttoo: A Thin and Feel-Through Tattoo for On-Skin Tactile Output. In Proc of the 31st Annual ACM Symp. on User Interface Soft. and Tech. (UIST '18), 365-378.

[73] X. Yi, C. Yu, W. Xu, X. Bi and Y. Shi. 2017. COMPASS: Rotational Keyboard on Non-Touch Smartwatches. CHI '17, 705-715.

[74] C. Yu, K. Sun, M. Zhong, X. Li, P. Zhao and Y. Shi. 2016. One-Dimensional Handwriting: Inputting Letters and Words on Smart Glasses. CHI '16, 71-82.

[75] C. Zhang, A. Bedri, G. Reyes, B. Bercik, O. T. Inan, T. E. Starner and G. D. Abowd. 2016. TapSkin: Recognizing On-Skin Input for Smartwatches. In Proc of the 2016 ACM Inter. Conf. on Interactive Surfaces and Spaces (ISS '16), 13-22.

[76] S. Zhu, T. Luo, X. Bi and S. Zhai. 2018. Typing on an Invisible Keyboard. CHI '18, 1-13.

[77] S. Ahn, S. Heo and G. Lee. 2017. Typing on a Smartwatch for Smart Glasses. In Proc of the 2017 ACM Inter. Conf. on Interactive Surfaces and Spaces (ISS '17), 201-209.

[78] J. Gong, Y. Zhang, X. Zhou and X. D. Yang. 2017. Pyro: Thumb-Tip Gesture Recognition Using Pyroelectric Infrared Sensing Proc of the 30th Annual ACM Symp. on User Interface Soft. and Tech. (UIST '17), 553-563.

[79] A. Gupta and R. Balakrishnan. 2016. DualKey: Miniature Screen Text Entry via Finger Identification. CHI '16, 59-70.

[80] H. L. Kao, A. Dementyev, J. A. Paradiso and C. Schmandt. 2015. NailO: Fingernails as an Input Surface. CHI '15, 3015-3018.

[81] H. L. Kao, C. Holz, A. Roseway, A. Calvo and C. Schmandt. 2016. DuoSkin: rapidly prototyping on-skin user interfaces using skin-friendly materials. In Proc of the 2016 ACM Inter. Symp. on Wearable Computers (ISWC '16), 16-23.

[82] I. S. MacKenzie, R. W. Soukoreff and J, Helga. 2011. 1 thumb, 4 buttons, 20 words per minute: Design and evaluation of H4-Writer. In Proc of the 24th annual ACM Symp. on User interface Soft. and Tech. (UIST '11), 471-480.

[83] A. Mottelson, C. Larsen, M. Lyderik, P. Strohmeier and J. Knibbe. 2016. Invisiboard: maximizing display and input space with a full screen text entry method for smartwatches. In Proc of the 18th Inter. Conf. on HCI w. Mobile Devices and Services (MobileHCI '16), 53-59.

[84] J. Lien, N Gillian, M. E. Karagozler, P. Amihood, C. Schwesig, E. Olson, H. R. and I. Poupyrev. 2016. Soli: ubiquitous gesture sensing with millimeter wave radar. ACM Trans. Graph. 5, 4, Art. 142 (July 2016), 19 pages.

[85] J. Lo, D. J. L. Lee, N Wong, D. Bui and E. Paulos. 2016. Skintillates: Designing and Creating Epidermal Interactions. In Proc of the 2016 ACM Conf. on Designing Interactive Systems (DIS '16), 853-864.

[86] C. Loclair, S. Gustafson and P. Baudisch. 2010. PinchWatch: a wearable device for one-handed microinteractions. In Proc. MobileHCI.

[87] M. Kytö, B. Ens, T. Piumsomboon, G. A. Lee, and M. Billinghurst. 2018. Pinpointing: Precise Head-and Eye-Based Target Selection for Augmented Reality. CHI '18. ACM NY, USA, Art. 81, 14 pages.

[88] V. Rajanna and J. P. Hansen. 2018. Gaze Typing in Virtual Reality: Impact of Keyboard Design, Selection Method, and Motion. In Proc of the 2018 ACM Symp. on Eye Tracking Research Applications (ETRA '18). ACM NY, USA, Art. 15, 10 pages.

[89] S. Ahn and G. Lee. 2019. Gaze-Assisted Typing for Smart Glasses. In Proc of the 32nd Annual ACM Symp. on User Interface Soft. and Tech. (UIST '19). ACM NY, USA, 857–869.

[90] Y. Zhang, W. Kienzle, Y. Ma, S. S. Ng, H. Benko, and C. Harrison. 2019. ActiTouch: Robust Touch Detection for On-Skin AR/VR Interfaces. In Proc of the 32nd Annual ACM Symp. on User Interface Soft. and Tech. (UIST '19). ACM NY, USA, 1151–1159.

[91] T. Hachisu, B. Bourreau, and K. Suzuki. 2019. EnhancedTouchX: Smart Bracelets for Augmenting Interpersonal Touch Interactions. CHI '19. ACM NY, USA, Paper 321, 12 pages.

[92] K. Suzuki, T. Hachisu, and K. Iida. 2016. EnhancedTouch: A Smart Bracelet for Enhancing Human-Human Physical Touch. CHI '16. ACM NY, USA, 1282-1293.

[93] V. Varga, M. Wyss, G. Vakulya, A. Sample, and T. R. Gross. 2018. Designing Groundless Body Channel Communication Systems: Performance and Implications. In Proc of the 31st Annual ACM Symp. on User Interface Soft. and Tech. (UIST '18). ACM NY, USA, 683-695.

[94] R. Xiao, J. Schwarz, N. Throm, A. D. Wilson, and H. Benko. 2018. MRTouch: adding touch input to head-mounted mixed reality. IEEE transactions on visualization and computer graphics 24, no. 4 (2018): 1653-1660.

[95] R. Xiao, T. Cao, N. Guo, J. Zhuo, Y. Zhang, and C. Harrison. 2018. LumiWatch: On-Arm Projected Graphics and Touch Input. CHI '18. ACM NY, USA, Paper 95, 11 pages.

[96] G. Cohn, D. Morris, S. Patel, and D. Tan. 2012. Humantenna: using the body as an antenna for real-time whole-body interaction. CHI '12. ACM NY, USA, 1901–1910.

[97] M. Teyssier, G. Bailly, C. Pelachaud, E. Lecolinet, A. Conn, and A. Roudaut. 2019. Skin-On Interfaces: A Bio-Driven Approach for Artificial Skin Design to Cover Interactive Devices. In Proc of the 32nd Annual ACM Symp. on User Interface Soft. and Tech. (UIST '19). ACM NY, USA, 307–322.

[98] R. Boldu, S. Jain, J. P. Forero Cortes, H. Zhang, and S. Nanayakkara. 2019. M-Hair: Creating Novel Tactile Feedback by Augmenting the Body Hair to Respond to Magnetic Field. In Proc of the 32nd Annual ACM Symp. on User Interface Soft. and Tech. (UIST '19). ACM NY, USA, 323–328.

[99] K. Katsuragawa, J Wang, Z. Shan, N. Ouyang, O. Abari, and D. Vogel. 2019. Tip-Tap: Battery-free Discrete 2D Fingertip Input. In Proc of the 32nd Annual ACM Symp. on User Interface Soft. and Tech. (UIST '19). ACM NY, USA, 1045–1057.

[100] Y. Gu, C. Yu, Z. Li, W. Li, S. Xu, X/ Wei, and Y. Shi. 2019. Accurate and Low-Latency Sensing of Touch Contact on Any Surface with Finger-Worn IMU Sensor. In Proc of the 32nd Annual ACM Symp. on User Interface Soft. and Tech. (UIST '19). ACM NY, USA, 1059–1070.

[101] X. Ma, E. Yang, and P. Fung. (2019). Exploring Perceived Emotional Intelligence of Personality-Driven Virtual Agents in Handling User Challenges. WWW'19.

[102] L. Clark, N. Pantidi, O. Cooney, P. Doyle, D. Garaialde, et al. 2019. What Makes a Good Conversation? Challenges in Designing Truly Conversational Agents. CHI '19. ACM NY, USA, Paper 475, 1–12.

[103] J. Robert. Moore and Raphael Arar. 2019. Conversational UX Design: A Practitioner's Guide to the Natural Conversation Framework. ACM NY, USA.

[104] A. Danielescu and G. Christian. 2018. A Bot is Not a Polyglot: Designing Personalities for Multi-Lingual Conversational Agents. CHI EA '18. ACM NY, USA, Paper CS01, 1–9.

[105] C. Toxtli, et al. Understanding Chatbot-mediated Task Management. CHI '18. ACM NY, USA, Paper 58, 1–6.

[106] C.H. Li, K. Chen, and Y.J. Chang. 2019. When There is No Progress with a Task-Oriented Chatbot: A Conversation Analysis. In Proc of the 21st Inter. Conf. on HCI w. Mobile Devices and Services (MobileHCI '19). ACM NY, USA, Art. 59, 1–6.

[107] X. Wang, S.S. Sohn, and M. Kapadia. 2019. Towards a Conversational Interface for Authoring Intelligent Virtual Characters. In Proc of the 19th ACM Inter. Conf. on Intelligent Virtual Agents (IVA '19). ACM NY, USA, 127–129.

[108] S. Schaffer and N. Reithinger. 2019. Conversation is multimodal: thus conversational user interfaces should be as well. In Proc of the 1st Inter. Conf. on Conversational User Interfaces (CUI '19). ACM NY, USA, Art. 12, 1–3.

[109] R. Jacques, E. Gerber, et al. 2019 Conversational agents: Acting on the wave of research and development. CHI EA 2019.

[110] Q.V. Liao, M.M. Hussain, et al. (2018). All work and no play? Conversations with a question-and-answer chatbot in the wild. CHI 2018.

[111] E. Luger, and G. Rosner. 2017. Considering the privacy design issues arising from conversation as platform. In R. Leenes, et al. (Eds.), Data protection and privacy: The age of intelligent machines. Computers, Privacy and Data Protection (10), Hart

Publishing.

[112] E. Luger and A. Sellen. 2016. "Like having a really bad PA": The gulf between user expectation and experience of conversational agents. Proc. CHI 2016, 5286–5297.

[113] J. Cassell, T. Bickmore, M. Billinghurst, et al. 1999. Embodiment in conversational interfaces: Rea. CHI '99. ACM NY, USA, 520–527.

[114] P. Seipel, A. Stock, et al. 2019. Adopting conversational interfaces for exploring OSGi-based software architectures in augmented reality. In Proc of the 1st Inter. Workshop on Bots in Software Engineering (BotSE '19). IEEE Press, 20–21.

[115] P. Mavridis, et al. 2019. Chatterbox: Conversational Interfaces for Microtask Crowdsourcing. In Proc of the 27th ACM Conf. on User Modeling, Adaptation and Personalization (UMAP '19). ACM NY, USA, 243–251.

[116] L.C. Klopfenstein, et al. 2017. The Rise of Bots: A Survey of Conversational Interfaces, Patterns, and Paradigms. In Proc of the 2017 Conf. on Designing Interactive Systems (DIS '17). ACM NY, USA, 555–565.

[117] Y. Li, T. Yao, et al. 2016. Video ChatBot: Triggering Live Social Interactions by Automatic Video Commenting. In Proc of the 24th ACM Inter. Conf. on Multimedia (MM '16). ACM NY, USA, 757–758.

[118] A. Veglis and T.A. Maniou. 2019. Embedding a chatbot in a news article: design and implementation. In Proc of the 23rd Pan-Hellenic Conf. on Informatics (PCI '19). ACM NY, USA, 169–172.

[119] D.Y. Huang, et al. Digitspace: Designing thumb-to-fingers touch interfaces for one-handed and eyes-free interactions. CHI '16, pages 1526–1537 NY, USA, 2016. ACM.

[120] J. F. Hessam, et al. 2017. Towards optimization of mid-air gestures for in-vehicle interactions. In Proc of the 29th Australian Conf. on Computer-Human Interaction (OZCHI '17). ACM NY, USA, 126–134.

[121] S.H. Lee, et al. 2015. On-wheel finger gesture control for in-vehicle systems on central consoles. In Adjunct Proc of the 7th Inter. Conf. on Automotive User Interfaces and Interactive Vehicular Applications (AutomotiveUI '15). ACM NY, USA, 94–99.

[122] J. G. Gaspar, et al. 2015. Examining the interaction between timing and modality in forward collision warnings. In Proc of the 7th Inter. Conf. on Automotive User Interfaces and Interactive Vehicular Applications (AutomotiveUI '15). ACM NY, USA, 313–319.

[123] D. R. Large, et al. 2019. "It's small talk, jim, but not as we know it.": engendering trust through human-agent conversation in an autonomous, self-driving car. In Proc of the 1st Inter. Conf. on Conversational User Interfaces (CUI '19). ACM NY, USA, Art. 22, 1–7.

[124] H. Hwan, et al. 2016. Complexity Overloaded in Smart Car: How to Measure Complexity of In-vehicle Displays and Controls? In Adjunct Proc of the 8th Inter. Conf. on Automotive User Interfaces and Interactive Vehicular Applications (AutomotiveUI '16 Adjunct). ACM NY, USA, 81–86.

[125] Y. Chen, et al. 2016. Pactolus: A Method for Mid-Air Gesture Segmentation within EMG. CHI EA '16. ACM NY, USA, 1760–1765.

[126] L. Qian, A. Plopski, N. Navab and P. Kazandides. 2018. Restoring the Awareness in the Occluded Visual Field for Optical See-Through Head-Mounted Displays. in IEEE Transactions on Visualization and Computer Graphics, vol. 24, no. 11, pp. 2936-2946, Nov. 2018.

[127] H. Kim, C. Coutrix, and A. Roudaut. 2018. Morphees+: Studying Everyday Reconfigurable Objects for the Design and Taxonomy of Reconfigurable UIs. CHI '18. ACM NY, USA, Paper 619, 1–14.

[128] I. P. S. Qamar, R. Groh, D. Holman, and A. Roudaut. 2018. HCI meets Material Science: A Literature Review of Morphing Materials for the Design of Shape-Changing Interfaces. CHI '18. ACM NY, USA, Paper 374, 1–23.

[129] I. Qamar, R. Groh, D. Holman, and A. Roudaut. 2019. Bridging the gap between material science and human-computer interaction. interactions 26, 5 (August 2019), 64–69.

[130] L. Angelini, et al. 2015. Tangible Meets Gestural: Comparing and Blending Post-WIMP Interaction Paradigms. In Proc of the Ninth Inter. Conf. on Tangible, Embedded, and Embodied Interaction (TEI '15). ACM NY, USA, 473–476.

[131] C. Myers, et al. 2018. Patterns for How Users Overcome Obstacles in Voice User Interfaces. CHI '18. ACM NY, USA, Paper 6, 1–7.

[132] T. Horak, et al. 2018. When David Meets Goliath: Combining Smartwatches with a Large Vertical Display for Visual Data Exploration. CHI '18. ACM NY, USA, Paper 19, 1–13.

[133] S. Swaminathan, et al. 2019. Input, Output and Construction Methods for Custom Fabrication of Room-Scale Deployable Pneumatic Structures. Proc. ACM Interact. Mob. Wearable Ubiquitous Technol. 3, 2, Art. 62 (June 2019), 17 pages.

[134] K. Fujinami, M. Kosaka, and B. Indurkhya. 2018. Painting an Apple with an Apple: A Tangible Tabletop Interface for Painting with Physical Objects. Proc. ACM Interact. Mob. Wearable Ubiquitous Technol. 2, 4, Art. 162 (December 2018), 22 pages.

[135] C.Y. Wang, et al. 2015. PalmType: Using Palms as Keyboards for Smart Glasses. In Proc of the 17th Inter. Conf. on HCI w. Mobile Devices and Services (MobileHCI '15). ACM NY, USA, 153–160.

[136] K. Palin, et al. 2019. How do People Type on Mobile Devices? Observations from a Study with 37,000 Volunteers. In Proc of the 21st Inter. Conf. on HCI w. Mobile Devices and Services (MobileHCI '19). ACM NY, USA, Art. 9, 1–12.

[137] H.S. Yeo, et al. 2018. Exploring Tangible Interactions with Radar Sensing. Proc. ACM Interact. Mob. Wearable Ubiquitous Technol. 2, 4, Art. 200 (December 2018), 25 pages.

[138] B. Shneiderman. 2000. The limits of speech recognition. Commun. ACM 43, 9 (2000), 63–65.

[139] M. X. Zhou, et al. 2019. Trusting Virtual Agents: The Effect of Personality. ACM Trans. Interact. Intell. Syst. 9, 2–3, Art. 10 (March 2019), 36 pages.

[140] J.I. Koh, et al. 2019. Developing a Hand Gesture Recognition System for Mapping Symbolic Hand Gestures to Analogous Emojis in Computer-Mediated Communication. ACM Trans. Interact. Intell. Syst. 9, 1, Art. 6 (March 2019), 35 pages.

[141] A. L. Baker, et al. 2018. Toward an Understanding of Trust Repair in Human-Robot Interaction: Current Research and Future Directions. ACM Trans. Interact. Intell. Syst. 8, 4, Art. 30 (November 2018), 30 pages.

[142] C. L. Sidner, et al. 2018. Creating New Technologies for Companionable Agents to Support Isolated Older Adults. ACM Trans. Interact. Intell. Syst. 8, 3, Art. 17 (July 2018), 27 pages.

[143] A. Paiva, et al. 2017. Empathy in Virtual Agents and Robots: A Survey. ACM Trans. Interact. Intell. Syst. 7, 3, Art. 11 (September 2017), 40 pages.

[144] T. Zhang, et al. 2016. The Effect of Embodied Interaction in Visual-Spatial Navigation. ACM Trans. Interact. Intell. Syst. 7, 1, Art. 3 (December 2016), 36 pages.

[145] Evelien van de Garde-Perik, et al. 2013. An analysis of input-output relations in interaction with smart tangible objects. ACM Trans. Interact. Intell. Syst. 3, 2, Art. 9 (August 2013), 20 pages.

[146] B. Kveton and S. Berkovsky. 2016. Minimal Interaction Content Discovery in Recommender Systems. ACM Trans. Interact. Intell. Syst. 6, 2, Art. 15 (July 2016), 25 pages.

[147] F. S. Parizi, et al. 2019. AuraRing: Precise Electromagnetic Finger Tracking. Proc. ACM Interact. Mob. Wearable Ubiquitous Technol. 3, 4, Art. 150 (December 2019), 28 pages.

[148] I. Belkacem, et al. (2019) TEXTile: Eyes-Free Text Input on Smart Glasses Using Touch Enabled Textile on the Forearm. INTERACT 2019. Lecture Notes in Computer Science, vol 11747. Springer, Cham.

[149] L. Pandey and A.S. Arif. 2019. Context-sensitive app prediction on the suggestion bar of a mobile keyboard. In Proc of the 18th Inter. Conf. on Mobile and Ubiquitous Multimedia (MUM '19). ACM NY, USA, Art. 45, 1–5.

[150] Z. Yang, et al. 2019. Investigating Gesture Typing for Indirect Touch. Proc. ACM Interact. Mob. Wearable Ubiquitous Technol. 3, 3, Art. 117 (September 2019), 22 pages.

[151] C. Park, et al. 2019. HandPoseMenu: Hand Posture-Based Virtual Menus for Changing Interaction Mode in 3D Space. In Proc of the 2019 ACM Inter. Conf. on Interactive Surfaces and Spaces (ISS '19). ACM NY, USA, 361–366.

[152] Z. H. Lim and P. O. Kristensson. 2019. An Evaluation of Discrete and Continuous Mid-Air Loop and Marking Menu Selection in Optical See-Through HMDs. In Proc of the 21st Inter. Conf. on HCI w. Mobile Devices and Services (MobileHCI '19). ACM NY, USA, Art. 16, 1–10.

[153] M. Landau and S. V. Delden. 2017. A System Architecture for Hands-Free UAV Drone Control Using Intuitive Voice Commands. In Proc of the Companion of the 2017 ACM/IEEE Inter. Conf. on Human-Robot Interaction (HRI '17). ACM NY, USA, 181–182.

[154] I. Wang, J. Smith, and J. Ruiz. 2019. Exploring Virtual Agents for Augmented Reality. CHI '19. ACM NY, USA, Paper 281, 1–12.

[155] D. Coyle, J. Moore, et al. (2012) I did that! Measuring users' experience of agency in their own actions. ACM CHI 2014. 2025-34.

[156] H. Limerick, J. W. Moore, and D. Coyle. 2015. Empirical Evidence for a Diminished Sense of Agency in Speech Interfaces. CHI '15. ACM NY, USA, 3967–3970.

[157] H. Kristina. 2018. Designing with the Body: Somaesthetic Interaction Design. CHIRA.

[158] I. Lee, S. Kim, and B. Lee. 2019. Geometrically Compensating Effect of End-to-End Latency in Moving-Target Selection Games. CHI '19. ACM NY, USA, Paper 560, 1–12.

[159] A.R. Lingley, et al. 2011. 'A single-pixel wireless contact lens display', J. Micromechanics and Microengineering, vol. 21, no. 12, 125014, pp. 1-8.

[160] TechCrunch Video, 2020, Mojo Vision is developing AR contact lenses. https://techcrunch.com/video-article/mojo-vision-is-developing-ar-contact-lenses/ (Access on January 17, 2020)

[161] L. Maria, et al. 2019. Foraging-based optimization of pervasive displays, Pervasive and Mobile Computing, Volume 55, 2019, Pages 45-58, ISSN 1574-1192.

[162] R. Delwo. 2018. What AR Field of View Feels Like. https://medium.com/@RDelly/what-ar-field-of-viewfeels-like-ecadd16afc5e, 2018.

[163] X. Qiao, et al. 2019. Web AR: A Promising Future for Mobile Augmented Reality—State of the Art, Challenges, and Insights. In Proc. of the IEEE, vol. 107, no. 4, pp. 651-666, April 2019.

[164] J. L. Derby, et al. 2019. Text Input Performance with a Mixed Reality Head-Mounted Display (HMD). Proc. of the Human Factors and Ergonomics Society Annual Meeting, 63(1), 1476–1480.

[165] C. Merenda, et al. 2019. Effects of "Real-World" Visual Fidelity on AR Interface Assessment: A Case Study Using AR Head-up Display Graphics in Driving. 2019 IEEE Inter. Symp. on Mixed and Augmented Reality (ISMAR), Beijing, China, 2019, pp. 145-156.

[166] N. Norouzi, et al. 2019. Walking Your Virtual Dog: Analysis of Awareness and Proxemics with Simulated Support Animals in Augmented Reality. 2019 IEEE Inter. Symp. on Mixed and Augmented Reality (ISMAR), Beijing, China, 2019, pp. 157-168.

[167] A. Marquardt, et al. 2019. Non-Visual Cues for View Management in Narrow Field of View Augmented Reality Displays. 2019 IEEE Inter. Symp. on Mixed and Augmented Reality (ISMAR), Beijing, China, 2019, pp. 190-201.

[168] H. Lee et al. 2019. Annotation vs. Virtual Tutor: Comparative Analysis on the Effectiveness of Visual Instructions in Immersive Virtual Reality. 2019 IEEE Inter. Symp. on Mixed and Augmented Reality (ISMAR), Beijing, China, 2019, pp. 318-327.

[169] A. Souchet, et al. 2019. Investigating Cyclical Stereoscopy Effects Over Visual Discomfort and Fatigue in Virtual Reality While Learning. 2019 IEEE Inter. Symp. on Mixed and Augmented Reality (ISMAR), Beijing, China, 2019, pp. 328-338.

[170] D.O. Faria, et al. 2019. Augmented reality head-up displays effect on drivers' spatial knowledge acquisition. Proc. of the Human Factors and Ergonomics Society Annual Meeting, 63(1), 1486–1487.

[171] J. D. Benedict, et al. 2019. The Intuitiveness of Gesture Control with a Mixed Reality Device. Proc. of the Human Factors and Ergonomics Society Annual Meeting, 63(1), 1435–1439.

[172] Y. Sun, et al. 2019. MagicHand: Interact with IoT Devices in Augmented Reality Environment. 2019 IEEE Conf. on Virtual Reality and 3D User Interfaces (VR), Osaka, Japan, 2019, pp. 1738-1743.

[173] Y. Park, et al. 2019. When IoT met Augmented Reality: Visualizing the Source of the Wireless Signal in AR View. In Proc. of the 17th Annual Inter. Conf. on Mobile Systems, Applications, and Services (MobiSys '19). ACM NY, USA, 117–129.

[174] D. Jo, and G. J. Kim. 2019. AR Enabled IoT for a Smart and Interactive Environment: A Survey and Future Directions. Sensors (Basel, Switzerland), 19(19), 4330.

[175] K. Michalakis, J. Aliprantis and G. Caridakis. 2018. Visualizing the Internet of Things: Naturalizing Human-Computer Interaction by Incorporating AR Features. In IEEE Consumer Electronics Magazine, vol. 7, no. 3, pp. 64-72, May 2018.

[176] S. Seneviratne, et al. 2017. A Survey of Wearable Devices and Challenges. IEEE Communications Surveys Tutorials. PP. 1-1.

[177] Y. Arifin, et al. 2018. User Experience Metric for Augmented Reality Application: A Review, Procedia Computer Science, Volume 135, 2018, Pages 648-656, ISSN 1877-0509,

[178] S. Anwar, et al. 2019. Hand Gesture Recognition: A Survey. In: Nath V., Mandal J. (eds) Nanoelectronics, Circuits and Communication Systems. Lecture Notes in Electrical Engineering, vol 511. Springer, Singapore

[179] A Hamacher, et al. 2019. Augmented Reality User Interface Evaluation – Performance Measurement of Hololens, Moverio and Mouse Input. Inter. Assoc. of Online Engineering. Retrieved January 18, 2020 from https://www.learntechlib.org/p/208272/.

[180] S., Nirjon, et al. 2015. TypingRing: A Wearable Ring Platform for Text Input. MobiSys '15.

[181] M. Zhong, et al. 2018. ForceBoard: Subtle Text Entry Leveraging Pressure. CHI '18. ACM NY, USA, Paper 528, 1–10.

[182] A. Gupta, et al. 2019. RotoSwype: Word-Gesture Typing using a Ring. CHI '19.

[183] J. Kim, et al. 2018. ThumbText: Text Entry for Wearable Devices Using a Miniature Ring. Graphics Interface.

[184] S. Ruan, et al. 2018. Comparing Speech and Keyboard Text Entry for Short Messages in Two Languages on Touchscreen Phones. Proc. ACM Interact. Mob. Wearable Ubiquitous Technol. 1, 4, Art. 159 (January 2018), 23 pages.

[185] S. Seneviratne, et al. 2017. A Survey of Wearable Devices and Challenges. In IEEE Communications Surveys Tutorials, vol. 19, no. 4, pp. 2573-2620, Fourthquarter.

[186] R.H. Liang, et al. 2019. InDexMo: exploring finger-worn RFID motion tracking for activity recognition on tagged objects. In Proc. of the 23rd Inter. Symp. on Wearable Computers (ISWC '19). ACM NY, USA, 129–134.

[187] J Orlosky, et al. 2013. Dynamic text management for see-through wearable and heads-up display systems. Inter. Conf. on Intelligent User Interfaces, Proc. IUI. 363-370.

[188] V. Kaptelinin and B. Nardi. 2012. Affordances in HCI: toward a mediated action perspective. CHI '12. ACM NY, USA, 967–976.

[189] W. Zhang, B. Han, and P. Hui. 2017. On the Networking Challenges of Mobile Augmented Reality. In Proc. of ACM SIGCOMM 2017 Workshop on Virtual Reality and Augmented Reality Network (VR/AR Network 2017), LA USA, August 2017.

[190] C. Bermejo, Z. Huang, T. Braud and P. Hui. 2017. When Augmented Reality meets Big Data. In Proc. of IEEE 37th Inter. Conf. on Distributed Computing Systems Workshops (ICDCSW), Atlanta, GA, 2017, pp. 169-174.

[191] S. Mayer, et al. 2020. Enhancing Mobile Voice Assistants with WorldGaze.Proc. of the 2020 CHI Conf. on Human Factors in Comp. Sys., ACM NY, USA

[192] Z. Xu, et al. 2020. BiTipText: Bimanual Eyes-Free Text Entry on a Fingertip Keyboard. CHI '20. ACM NY, USA, 1–13.

[193] F. Zhu and T. Grossman. 2020. BISHARE: Exploring Bidirectional Interactions Between Smartphones and Head-Mounted Augmented Reality. CHI '20. ACM NY, USA, 1–14.

[194] N. Mohammadi and J. E. Taylor. 2017. Smart city digital twins. 2017 IEEE Symp. Series on Computational Intelligence (SSCI), Honolulu, HI, 2017, pp. 1-5.

[195] H. S. Alavi, et al. 2019. Introduction to Human-Building Interaction (HBI): Interfacing HCI with Architecture and Urban Design. ACM Trans. Comput.-Hum. Interact. 26, 2, Art. 6 (March 2019), 10 pages.

[196] A. Henrysson and M. Ollila. 2004. UMAR: Ubiquitous Mobile Augmented Reality. In Proc. of the 3rd Inter. Conf. on Mobile and ubiquitous multimedia (MUM '04). ACM NY, USA, 41–45.

[197] A. Morrison, et al. 2009. Like bees around the hive: a comparative study of a mobile augmented reality map. CHI '09. ACM NY, USA, 1889–1898.

[198] U. Rashid, et al. 2012. Factors influencing visual attention switch in multi-display user interfaces: a survey. In Proc. of the 2012 Inter. Symp. on Pervasive Displays (PerDis '12). ACM NY, USA, Art. 1, 1–6.

[199] Y.P. Yau, L.H. Lee, et al. 2020. How Subtle Can It Get? A Trimodal Study of Ring-sized Interfaces for One-Handed Drone Control. Proc. ACM Interact. Mob. Wearable Ubiquitous Technol. 4, 2, Art. 63 (June 2020), 29 pages.

[200] N. Dayama, et al. 2020. GRIDS: Interactive Layout Design with Integer Programming. CHI '20.

[201] T. Nam and T. A. Pardo. 2011. Conceptualizing smart city with dimensions of technology, people, and institutions. In Proc. of the 12th Annual Inter. Digital Government Research Conference: Digital Government Innovation in Challenging Times (dg.o '11). ACM NY, USA, 282–291.

[202] T. Braud, P. Zhou, J. Kangasharju  P. Hui. 2020. Multipath Computation Offloading for Mobile Augmented Reality. 2020 IEEE Inter. Conf. on Pervasive Computing and Communications (PerCom), pp. 1-10.

[203] P. Zhou, W. Zhang, T. Braud, P. Hui and J. Kangasharju. 2019. Enhanced Augmented Reality Applications in Vehicle-to-Edge Networks. 2019 22nd Conf. on Innovation in Clouds, Internet and Networks and Workshops (ICIN), Paris, France, 2019, pp. 167-174.

[204] P. Zhou, W. Zhang, T. Braud, P. Hui, and J. Kangasharju. 2018. ARVE: Augmented Reality Applications in Vehicle to Edge Networks. In Proc. of the 2018 Workshop on Mobile Edge Communications (MECOMM'18). ACM NY, USA, 25–30.

[205] A. Luusua, J. Ylipulli, et al. 2015. Evaluation Probes. CHI '15. Assoc. for Comp. Mach., NY. USA, 85–94.

[206] S. Hollister. 2020. Google Glass is adding Meet so remote supervisors can see through field workers' eyes. Retrieved March 15, 2021, from https://www.theverge.com/2020/10/14/21516402/google-glass-meet-beta-enterprise-edition-remote-work-skype-microsoft-hololens

[207] J. Bailenson. 2018. Virtual reality can help make people more empathetic. Retrieved March 15, 2021, from https://news.stanford.edu/2018/10/17/virtual-reality-can-help-make-people-empathetic/

[208] T. Piumsomboon, et al. 2017. Empathic Mixed Reality: Sharing What You Feel and Interacting with What You See. 2017 Inter. Symp. on Ubiquitous Virtual Reality (ISUVR), Nara, Japan, 2017, pp. 38-41.

[209] A. Coplan. 2011. Will the Real Empathy Please Stand Up? A Case for a Narrow Conceptualization. The Southern J. Philosophy. 49.

[210] VR APP "I Have Low Vision". 2017. Retrieved from https://www.tengobajavision.com/en/app/vr-simulator/

[211] B. Spanlang, et al. 2014. How to build an embodiment lab: achieving body representation illusions in virtual reality. Front. Robot. AI 1:9.

[212] T. Wang, et al. 2020. CAPturAR: An Augmented Reality Tool for Authoring Human-Involved Context-Aware Applications. In Proc. of the 33rd Annual ACM Symp. on User Interface Soft. and Tech. (UIST '20). ACM NY, USA, 328–341.

[213] J. Kim, and V. Interrante. 2017. Dwarf or Giant: The Influence of Interpupillary Distance and Eye Height on Size Perception in Virtual Environments. ICAT-EGVE 2017.

[214] M. Pouke, K. J. Mimnaugh, T. Ojala and S. M. LaValle, 2020. The Plausibility Paradox For Scaled-Down Users In Virtual Environments. 2020 IEEE Conf. on Virtual Reality and 3D User Interfaces (VR), Atlanta, GA, USA, 2020, pp. 913-921.

[215] J. Nishida, et al. 2020. HandMorph: a Passive Exoskeleton that Miniaturizes Grasp. In Proc. of the 33rd Annual ACM Symp. on User Interface Soft. and Tech.. ACM NY, USA, 565–578.

[216] S. Pratte, A. Tang, and L. Oehlberg. 2021. Evoking Empathy: A Framework for Describing Empathy Tools. In Proc. of the Fifteenth Inter. Conf. on Tangible, Embedded, and Embodied Interaction (TEI '21). ACM NY, USA, Art. 25, 1–15.

[217] T. Feuchtner and J. Müller. 2017. Extending the Body for Interaction with Reality. CHI '17. ACM NY, USA, 5145–5157.

[218] C. Klocke. 2018. I am a Man: Vr civil RIGHTS APP. Retrieved March 16, 2021, from https://design.ncsu.edu/blog/2018/01/25/i-am-a-man-vr-civil-rights-app/

[219] C.D. Cogburn, et al. 2018. 1000 cut journey. ACM SIGGRAPH 2018 Virtual, Augmented, and Mixed Reality.

[220] Z. Wu, et al. 2020. Towards Detecting Need for Empathetic Response in Motivational Interviewing. In Companion Publication of the 2020 Inter. Conf. on Multimodal Interaction (ICMI '20 Companion). ACM NY, USA, 497–502.

[221] C.L. Bennett and D. K. Rosner. 2019. The Promise of Empathy: Design, Disability, and Knowing the "Other". CHI '19. ACM NY, USA, Paper 298, 1–13.

[222] J. Hong and L. Findlater. 2018. Identifying Speech Input Errors Through Audio-Only Interaction. CHI '18. ACM NY, USA, Paper 567, 1–12.

[223] S. Feiner, et al. 1993. Windows on the world: 2D windows for 3D augmented reality. In Proc. of the 6th annual ACM Symp. on User interface Soft. and Tech. (UIST '93). ACM NY, USA, 145–155.

[224] S. Feiner, et al. 1997. A touring machine: Prototyping 3D mobile augmented reality systems for exploring the urban environment. Personal Technologies 1, 208–217.

[225] T. Langlotz, et al. 2012. Sketching up the world: in situ authoring for mobile Augmented Reality. Personal Ubiquitous Comput. 16, 6 (August 2012), 623–630.

[226] T. Langlotz, et al. 2011. Robust detection and tracking of annotations for outdoor augmented reality browsing, Computers and Graphics, Volume 35, Issue 4, 2011, Pages 831-840, ISSN 0097-8493.

[227] B. MacIntyre, E. M. Coelho and S. J. Julier. 2002. Estimating and adapting to registration errors in augmented reality systems. Proc. IEEE Virtual Reality 2002, Orlando, FL, USA, pp. 73-80.

[228] B. MacIntyre and E. Machado Coelho. 2000. Adapting to dynamic registration errors using level of error (LOE) filtering. Proc. IEEE and ACM Inter. Symp. on Augmented Reality (ISAR 2000), Munich, Germany, pp. 85-88.

[229] J. S. Pierce and R. Pausch. 2002. Comparing voodoo dolls and HOMER: exploring the importance of feedback in virtual environments. CHI '02. ACM NY, USA, 105–112.

[230] S. Jeffrey, et al. 1999. Voodoo dolls: seamless interaction at multiple scales in virtual environments. In Proc. of the 1999 Symp. on Interactive 3D graphics (I3D '99). ACM NY, USA, 141–145.

[231] D. Schmalstieg and T. Höllerer. Augmented Reality - Principles and Practice. Addison-Wesley Professional, 2016.

[232] J. LaViola, et al. 3D User Interfaces: Theory and Practice. Addison Wesley, 2017.

[233] G. Kipper and J. Rampolla. 2012. Augmented Reality: An Emerging Technologies Guide to AR (1st. ed.). Syngress Publishing.

[234] J. S. Roo, et al. 2017. Inner Garden: Connecting Inner States to a Mixed Reality Sandbox for Mindfulness. CHI '17. ACM NY, USA, 1459–1470.

[235] C. Xie, et al. 2016. Large Scale Interactive AR Display Based on a Projector-Camera System. In Proc. of the 2016 Symp. on Spatial User Interaction (SUI '16). ACM NY, USA, 179.

[236] J. Hartmann, et al. 2020. AAR: Augmenting a Wearable Augmented Reality Display with an Actuated Head-Mounted Projector. In Proc. of the 33rd Annual ACM Symp. on User Interface Soft. and Tech. (UIST '20). ACM NY, USA, 445–458.

[237] H. Aoki and J. Rekimoto. 2019. Extramission: A Large Scale Interactive Virtual Environment Using Head Mounted Projectors and Retro-reflectors. In Symp. on Spatial User Interaction (SUI '19). ACM NY, USA, Art. 8, 1–9.

[238] K. Č. Pucihar, et al. 2014. The use of surrounding visual context in handheld AR: device vs. user perspective rendering. CHI '14. ACM NY, USA, 197–206.

[239] P. Wacker, et al. 2020. Heatmaps, Shadows, Bubbles, Rays: Comparing Mid-Air Pen Position Visualizations in Handheld AR. CHI '20. ACM NY, USA, 1–11.

[240] n.d. Samsung is racing to beat Apple to market WITH AR/AV SMARTGLASSES that uses snap on prescription lenses. Retrieved March 20, 2021, from https://www.patentlyapple.com/patently-apple/2021/03/samsung-is-racing-to-beat-apple-to-market-with-arav-smartglasses-that-uses-snap-on-prescription-lenses.html/

[241] Urban Archive, I. (2017, March 07). Urban archive EXPLORE HISTORY. Retrieved March 23, 2021, from https://apps.apple.com/us/app/urban-archive-nyc/id1154228951

[242] Lens studio. (n.d.). Retrieved March 23, 2021, from https://lensstudio.snapchat.com/templates/landmarker/guide/

[243] 8th Wall. (n.d.). Retrieved March 23, 2021, from https://www.8thwall.com/

[244] Acute Art. Unreal City at Home. Retrieved March 23, 2021, from https://acuteart.com/artist/unreal-city/

[245] S. Stein. Niantic is testing Pokemon Go on HoloLens 2 to let you play with friends remotely. Retrieved March 23, 2021, from https://www.cnet.com/news/niantic-is-testing-pokemon-go-on-hololens-2-to-let-play-with-friends-remotely/

[246] S. Stein. What it's like to play Super Mario in AR, and why it matters: Mario on a HoloLens is weird and tiring – and it's also the future. Retrieved March 23, 2021, from https://www.cnet.com/news/what-its-like-to-play-super-mario-in-ar-and-why-it-matters/

[247] C. Coward. Play Mario Kart IRL with HoloLens: Ian Charnas has managed to use Microsoft HoloLens augmented reality headsets to bring Mario Kart to life. Retrieved March 23, 2021, from https://www.hackster.io/news/play-mario-kart-irl-with-hololens-1ea3a136d436/

[248] Immersal SDK. (n.d.). Retrieved March 23, 2021, from https://immersal.com/

[249] F. GERMAIN. AR house : Groundplane, Model Target   Area Target. Retrieved March 23, 2021, from https://www.youtube.com/watch?v=rViYSIn-NOg/

[250] Google Developers. What's new in AR (Google I/O '18). Retrieved March 23, 2021, from https://www.youtube.com/watch?v=MeZcQguH124/

[251] Google Developers. What's new in AR (Google I/O '19). Retrieved March 23, 2021, from https://www.youtube.com/watch?v=TQSaPsKHPqs/

[252] L. Zhang, et al. (2018). Visualizing Toronto City Data with HoloLens: Using Augmented Reality for a City Model. IEEE Consumer Electronics Magazine. 7. 73-80.

[253] J. M. Jensen, et al. 2018. Informing Informal Caregivers About Dementia Through an Experience-Based Virtual Reality Game. In Conf. on Smart Learning Ecosystems and Regional Development, pp. 125-132, Springer, Cham.

[254] M. Koelle, et al. 2015. Don't look at me that way! Understanding User Attitudes Towards Data Glasses Usage. In Proc. of the 17th Inter. Conf. on HCI w. Mobile Devices and Services (MobileHCI '15). ACM NY, USA, 362–372.

[255] M. Koelle, et al. 2017. All about Acceptability? Identifying Factors for the Adoption of Data Glasses. CHI '17. ACM NY, USA, 295–300.

[256] M. Koelle, et al. 2020. Social Acceptability in HCI: A Survey of Methods, Measures, and Design Strategies. CHI '20. ACM NY, USA, 1–19.

[257] N. Kelly, and S. Gilbert. 2016. The WEAR scale: Developing a measure of the social acceptability of a wearable device. CHI EA 2016. 2864—2871.

[258] I. Poupyrev, et al. 2016. Project Jacquard: Interactive Digital Textiles at Scale. CHI '16. ACM NY, USA, 4216–4227.

[259] D. Dobbelstein, et al. 2017. PocketThumb: a Wearable Dual-Sided Touch Interface for Cursor-based Control of Smart-Eyewear. Proc. ACM Interact. Mob. Wearable Ubiquitous Technol. 1, 2, Art. 9 (June 2017), 17 pages.

[260] K. Klamka and R. Dachselt. 2018. ARCord: Visually Augmented Interactive Cords for Mobile Interaction. CHI EA '18. ACM NY, USA, Paper LBW623, 1–6.

# APPENDIX

Table 1. Issues of AR Interface design.

| Issue(s) | Purpose(s) of the study |
|---|---|
| Cognition (Dual-task) | Studying the effect of content placement to user cognitive loads in dual-task situation [20] |
| | Studying the effect of mobile situations (climbing [36], vehicle [165], unexpected event [38]) in dual-task situation |
| | Studying the effect of content placement to user noticability in dual tasks [27] |
| Content Access | Proposing the display foraging theory to optimize content placement under the premise of limited user attention and read time [161] |
| | An adaptive framework to organize information for small-size display in lean and user-centric manners [6] |
| | Studying the effectiveness of recommendation systems reducing the cost of interaction (e.g., # of clicks) [146] |
| | Designing shortcut icons governing by in-context recommendation system on top of the workbench to reduce task time [149] |
| Field of View | Studying the effect of limited FOV to user task performance [26] |
| | Investigating the feasibility of alternative navigation clues with audio and tactile feedback to reduce visual clutters [167] |
| | Studying the effect of headset occlusion to the user response time to critical situations [126] |
| | Understanding screen sizes on various augmented reality smartglasses [162] |
| | A tool of compensating the loss of FOV by providing an overview window on the remaining FOV [19] |
| Human Vision | The effect of wearable device prism position to peripheral visual functions [29] |
| | Designing text entry approaches relying on peripheral vision [2] |
| | Investigating the position of navigation clues using peripheral vision [7] |
| | Studying the effect of text position and transformation in peripheral vision to reading efficiency [16] |
| Readability | Studying the effect of text style, colour, illuminance to the user readability [49] |
| | Proposing a text presentation method in various mobile scenarios (walking, standing, crowd, stair) [13] |
| | Proposing background detection approaches to improve the viewability of objects [28] and text [187] |
| | Studying the effect of VR object sizes to optical illusion [50] |
| | Studying the effect of bar chart presentation to the Ebbinghaus illusion [15] |
| | Studying the effect of font design under sharkness condition to user readability [12] |
| | Comparing the text reading speed and accuracy in two walking paths (with/without obstacles) [18] |
| | Studying the effect of font size and style to readability [24] |
| | Studying the effect of text position and presentation in walking and sitting postures to readability [21] |
| Social (Interruptions) | Investigating the effect of notification placement to the information noticability and user intrusiveness [20] |
| | Understanding the bystander perception about the time to interupt the headset users [47] |

Table 2. Core research areas and related research problems to tackle.

| | Four Areas | | | | Sample Research Problems |
|---|---|---|---|---|---|
| Networking | Output | Input | CUIs | | |
| x | x | | | | How to compensate for the impact of networking latency to reserve user cognition loads in dual-task scenarios. |
| x | x | | | | What is the user perceptions to the granularity level of 3D objects under various networking latency (Cloud vs Edge servers) |
| x | | x | | | How to alleviate the mistargeting in pointing-and-selection task due to unavoidable networking latency? |
| x | | x | | | What are networking requirements to make satisfied user interaction with various gestural types and interaction techniques? |
| x | | | x | | What appearance of the virtual agent can enhance the user tolerance to degraded performance due to unreliable network? |
| x | | | x | | How to handle context-mismatch in the response of CUIs due to high networking latency? |
| | x | x | | | What are the designs of text selection techniques, if the textual content is located at the peripheral vision? |
| | x | x | | | What are the text entry approaches leveraging peripheral vision? |
| | x | | x | | What is the font size and text style for chatroom style CUIs in mobile scenarios? (e.g., walking, stairs, crowds, etc.) |
| | x | | | | How to fit the agents and visual confirmations into the small-size screen real estate? |
| | | x | x | | What are the temporal models to switch between the on-body input techniques and the CUIs? |
| | | x | x | | Can we leverage the sensors on the on-body input devices to improve the intelligent of CUIs? (e.g., knowing the environmental context) |

Table 3. Input approaches: Category, Modality, Embodiment Strategy, Purposes.

| Reference | Category | Modality | Embodiment Strategy | Key Purpose(s) with AR |
|---|---|---|---|---|
| [83] | On-watch | Touch-sensitive surface | Common knowledge of alphabetical order, i.e., A – Z | Text input |
| [74] | On-glasses | Touch-sensitive surface | Common knowledge of stoke direction of alphabets | Text input |
| [2] | On-device | Computer vision (Headset) | Common knowledge of alphabetical order, i.e., A – Z | Text input |
| [76] | On-device | Touch-sensitive surface | Common knowledge of key positions in a full QWERTY keyboard | Text input |
| [64] | On-device | Touch-sensitive surface | Common knowledge of key positions in a full QWERTY keyboard | Text input |
| [151] | On-body | Computer vision (Headset) | Intuitive Hand Gestures mapped to menu projected on the palm and forearm | AR menu |
| [90] | On-body | Capacitive | Menus and buttons mapped to the plam and forearm, finger-to-forearm interaction | AR menu |
| [65] | On-body | Nil (Formative study) | The coexistence of tactile feedbacks and on-skin devices | Designing tactile feedbacks with epidermal devices |
| [140] | On-body | Computer vision (Headset) | Similarity between body gestures and emojis | Emojis input |
| [72] | On-body | Tactile | Alternative user perceptions other than visual and audio loads | Feel-through feedbacks on sensitive skin surface |
| [98] | On-body | Tactile | Alternative user perceptions other than visual and audio loads | Feel-through feedbacks on skin hair |
| [84] | On-body | Radar | Micro-gestures between fingers of one hand | Gesture-to-command |
| [78] | On-body | Infrared | Thumb-to-fingertip microgestures (e.g., circle, triangle, rub, etc.) | Gesture-to-command |
| [86] | On-body | Computer vision (Wrist-worn) | Micro-gestures between fingers of one hand | Gesture-to-command |
| [120] | On-body | Computer vision | Intuitive Hand Gestures for controlling in-vehicle interfaces | Gesture-to-command |
| [171] | On-body | Computer vision | Learning new hand gestures | Gesture-to-command |
| [71] | On-body | Electrodes and circuits | Touch on sensitive and spacious skin surface, accomodating various gestures | Gesture-to-command |
| [81] | On-body | Near-field Communication (NFC) | Enabling the spacious skin surface on human body as control devices | Gesture-to-command |
| [99] | On-body | RFID | Micro-gestures driven by thumb-to-fingertip interaction | Gesture-to-command |
| [91, 92] | On-body | Electrodes and circuits | Touch events between two hands of two people | In-city social events by Interpersonal Interaction |
| [69] | On-body | Fluidic material | Alternative user perceptions other than visual and audio loads | Information display and notifications |
| [85] | On-body | LED lights | The spacious skin surface becomes swift and easy-to-reach channels for notifications | Information display and notifications |
| [80] | On-body | Touch-sensitive surface | Miniature-size interface on a nail and fingertip-to-nail interactions | Interfaces mapped Press and swipe gestures |
| [70] | On-body | Buttons made of configurable material | Extended body part with augmented sensing of press, shear and pinch gestures | Interfaces mapped the gestures |
| [172] | On-body | Computer vision | Intuitive Hand Gestures mapped to menu projected on the IoT devices | IoT device interaction (e.g., switch on/off light bulbs) |
| [75] | On-body | (Tap-on-skin) Acoustic sound | Spacious Forearm for multi-button menu with finger-to-forearm touches | Multi-button on-skin menus |
| [53] | On-body | Touch-sensitive surface | Thumb-to-finger space interaction of two-handed space | Object selection and Text input |
| [3] | On-body | Computer vision (Headset) | Employing deterous fingertip to direct object manipulations | Point-and-select |
| [147] | On-body | Electromagnetic field | Employing deterous fingertip to direct object manipulations | Point-and-select |
| [100] | On-body | IMUs and Computer Vision (Headsets) | Employing deterous fingertip to direct object manipulations | Point-and-select |
| [52] | On-body | IMUs and haptics | Off-hand interaction for natural body posture and social acceptane | Point-and-select, scrolling, text entry |
| [60] | On-body | Touch-sensitive and Force-sensitive surface | Thumb-size button on size-constrained interfaces perhaps located on the index finger | Text acquisition |
| [4] | On-body | Touch-sensitive and Force-sensitive surface | Thumb-to-finger space interaction of one-handed space, the skill of force exertion | Text input |
| [61] | On-body | Touch-sensitive surface | Thumb-to-finger space interaction of one-handed space | Text input |
| [5] | On-body | Touch-sensitive surface | Thumb-to-fingertip interaction of one-finger space | Text input |
| [88] | On-body | Gaze | Gaze focuses on the region of interests, and works on the region accordingly | Text input |
| [79] | On-body | Optical | Each finger own own unique meanings on a touch surface | Text input |
| [95] | On-body | Computer vision (Wrist-worn) | Virtual overlays on the spacious forearm driven by finger-to-forearm touches | Virtual overlays on the forearm |
| [186] | Everyday Objects | IMUs and RFID | By knowing the adjacent objects, augmented sensing and reserving visual resources | Activity Recognition with tangible objects |
| [9] | Everyday Objects | Computer vision (Headset) | Intuitive Hand Gestures to control on-wall display and office utility e.g., printers | AR Office workspace |
| [97] | Everyday Objects | Artifical skins | Skin-alike texture on everyday objects | Gestural inputs on touch-sensitive and moldable surface |
| [63] | Everyday Objects | Nil (Formative study) | Micro-gestures between fingers of one hand holding a small item | Gesture design in mobile scenarios |
| [133] | Everyday Objects | (Ultrasonic) Acoustic sound | Surface of building infrastructures become sensible to human touches | Interactions with WIMP on building (e.g., wall) |
| [132] | Everyday Objects | Touch-sensitive surface | Gestural inputs on watch devices | Manipulate objects on large displays |
| [137] | Everyday Objects | Radar | By knowing the adjacent objects, augmented sensing and reserving visual resources | Object recognition |
| [134] | Everyday Objects | Infrared | Shape and Colour of tangible objects | Painting task |
| [127] | Everyday Objects | Nil (Formative study) | Adaptive, moldable, shape-changing, reconfigurable objects | Post-WIMP UI design on the objects |
| [145] | Everyday Objects | Nil (Formative study) | Human action cycle with smart tangible objects | Smart windows, connecting cards, colour messaging |